\documentclass [12pt]{article}
\usepackage{amsmath,amssymb,cite}
\usepackage{appendix}
\usepackage{feynmp}
\usepackage[vcentermath,enableskew]{youngtab}
\usepackage{tabularx}
\usepackage[english]{babel}
\usepackage[dvips]{graphicx}
\usepackage{indentfirst}
\usepackage{epsfig}
\usepackage[]{hyperref}
\usepackage[section]{placeins}
\usepackage[stable]{footmisc}
\usepackage{appendix}
\usepackage[english]{babel}
\usepackage[dvips]{graphicx}
\usepackage{indentfirst}
\usepackage{epsfig}
\usepackage{slashed}
\usepackage{fancyhdr}
\numberwithin{equation}{section}

\begin{document}

\title{
%\hspace{15cm}{\small DIAS-05-06}\\
Emergent fuzzy geometry and fuzzy physics in $4$ dimensions}
%Emergent Geometry, (Non)Commutative Gauge Theory, and Topology Change in $4$ Dimensions}

\author{Badis Ydri, Rouag Ahlam, Ramda Khaled\\
Department of Physics, Faculty of Sciences, Badji Mokhtar Annaba University,\\
 Annaba, Algeria.
}

\maketitle
\abstract{A detailed Monte Carlo calculation of the phase diagram of bosonic IKKT Yang-Mills matrix models in three and six dimensions with quartic mass deformations is given. Background emergent fuzzy geometries in two and four dimensions are observed with a fluctuation given by a noncommutative $U(1)$ gauge theory very weakly coupled to normal scalar fields. The geometry, which is determined dynamically, is given by the fuzzy spheres ${\bf S}^2_N$ and ${\bf S}^2_N\times{\bf S}^2_N$ respectively.  The three and six matrix models are in the same universality class with some differences. For example, in two dimensions the geometry is completely stable,  whereas in four dimensions the geometry is stable only in the limit $M\longrightarrow \infty$, where $M$ is the mass of the normal fluctuations. The behavior of the eigenvalue distribution in the two theories is also different. We also sketch how we can obtain a stable fuzzy four-sphere ${\bf S}^2_N\times{\bf S}^2_N$ in the large $N$ limit for all values of $M$ as well as models of topology change in which the transition between spheres of different dimensions is observed. The stable fuzzy spheres in two and four dimensions act precisely as regulators which is the original goal of fuzzy geometry and fuzzy physics. Fuzzy physics and fuzzy field theory on these spaces are briefly discussed.   }
\newpage
\tableofcontents
%\newpage
\section{Introduction}
The IKKT model is equivalent to Connes' approach to geometry!

A commutative/noncommutative space in Connes' approach to geometry is given in terms of a spectral triple $({\cal A},\Delta,{\cal H})$ rather than in terms of a set of points \cite{Connes:1996gi}. ${\cal A}$ is the algebra of functions or bounded operators on the space, $\Delta$  is the Laplace operator or, in the case of spinors, the Dirac operator, and ${\cal H}$ is the Hilbert space on which the algebra of bounded operators and the differential operator $\Delta$ are represented. 

In the IKKT model the geometry is in a precise sense emergent. And thus from this point of view it is obviously superior to Connes' noncommutative geometry. The algebra ${\cal A}$ is given, in the large $N$ limit, by Hermitian matrices with smooth eigenvalue distribution and bounded square trace \cite{Sochichiu:2000ud}. The Laplacian/Dirac operator is given in terms of the background solution while the Hilbert space ${\cal H}$ is given by the adjoint representation of the gauge group $U(N)$.

In this article we will study IKKT Yang-Mills matrix models with quartic mass deformations in three and six dimensions with $SO(3)$ and $SO(3)\times SO(3)$ symmetries which will lead naturally to the fuzzy two-sphere ${\bf S}^2_N$ and to the fuzzy four-sphere ${\bf S}^2_N\times{\bf S}^2_N$ respectively.

Noncommutative gauge theory on the fuzzy two-sphere \cite{Madore:1991bw,Hoppe:1982} was introduced in \cite{Iso:2001mg}. It was derived as the low energy dynamics of open strings moving in a background magnetic field with $S^3$ metric  in \cite{Alekseev:2000fd}. This theory consists of the Yang-Mills term which can be obtained from the reduction to zero dimensions of ordinary $U(N)$ Yang-Mills theory in $3$ dimensions and a Chern-Simons term due to Myers effect \cite{Myers:1999ps}. The model was studied perturbatively in \cite{CastroVillarreal:2004vh} and  \cite{Azuma:2004ie} and nonperturbatively in \cite{Azuma:2004zq}. This model contains beside the usual two-dimensional gauge field a scalar fluctuation normal  to the sphere. In \cite{O'Connor:2006wv} a generalized model was proposed and studied in which this normal scalar field was suppressed by giving it a potential with very large mass. This was studied further in \cite{DelgadilloBlando:2008vi,DelgadilloBlando:2007vx}  where the instability of the sphere was interpreted along the lines of an emergent geometry phenomena. 

In \cite{Steinacker:2003sd} an elegant random matrix model with a single matrix was shown to be equivalent to a gauge theory on the fuzzy sphere with a very particular form of the potential which in the large $N$ limit leads to a decoupled normal scalar  fluctuation. In \cite{Steinacker:2007iy,Steinacker:2007iq,Ydri:2007px,Ydri:2006xw} an alternative model of gauge theory on the fuzzy sphere was proposed in which field configurations live in the Grassmannian manifold $U(2N)/(U(N+1)\times U(N-1))$.  In \cite{Steinacker:2007iy,Steinacker:2007iq} this model was shown to possess the same partition function as the commutative model via the application of the powerful localization techniques.

Noncommutative gauge theory on the fuzzy four-sphere ${\bf S}^2_N\times{\bf S}^2_N$ which is given by a six matrix model with global $SO(3)\times SO(3)$ symmetry containing at most quartic powers of the matrices was proposed in \cite{DelgadilloBlando:2006dp}. The value $M=1/2$ of the mass deformation parameter corresponds to the model studied \cite{Behr:2005wp} which can also be shown to correspond to a random matrix model with two matrices. This theory involves two normal scalar fields plus a four-dimensional gauge field. Again the mass deformation parameter $M$ is essentially the mass of these normal fluctuations and thus for large $M$ these scalar fields become weakly coupled. In \cite{Ydri:2016kua} an interpretation of these normal scalar fields as dark energy as dark energy is put forward.

The main results of this article are as follows:
\begin{itemize}

\item

The dynamically emergent geometry, which is given by a fuzzy two-sphere ${\bf S}^2_N$,  in the $3-$dimensional IKKT matrix models, is found to be stable for all values of the deformation parameter $M$. The critical gauge coupling constant $\tilde{\alpha}$ is found to scale as in equation (\ref{stable_sphere}), i.e. as $\tilde{\alpha}\sim 1/\sqrt{N}$. The sphere-to-matrix transition line is pushed to 0 and only one phase survives.

\item 

The $6-$dimensional IKKT matrix model exhibits a phase transition from a geometrical phase at low temperature, given by a fuzzy four-sphere ${\bf S}^2_N\times{\bf S}^2_N$ background, to a Yang-Mills  matrix phase with no background geometrical structure at high temperature. The inverse temperature $\beta$ is here identified with the gauge coupling constant $\tilde{\alpha}$.

\item The transition is exotic in the sense that we observe, for small values of $M$, a discontinuous jump in the entropy, characteristic of a 1st order
transition, yet with divergent critical fluctuations and a divergent
specific heat with critical exponent $\alpha=1/2$. The critical gauge coupling constant is pushed downwards as the 
scalar field mass is increased.  

\item For small $M$, the system in the Yang-Mills phase is well
approximated by $6$ decoupled matrices with a joint eigenvalue distribution which is uniform inside a ball in ${\bf R}^6$. This gives what we call the $d=6$ law given by equation (\ref{d=6}). For large $M$, the transition from the four-sphere phase ${\bf S}^2_N\times{\bf S}^2_N$ to the Yang-Mills matrix phase turns into a crossover and the eigenvalue distribution in the Yang-Mills matrix phase changes from the $d=6$ law to a uniform distribution. 

\item  In the Yang-Mills matrix phase the specific heat is equal to $3/2$ which coincides with the specific heat of $6$ independent matrix models with quartic potential in the high temperature limit. Once the geometrical phase is well established the specific heat 
takes the value $5/2$ with the gauge field contributing $1/2$ and the two scalar fields each contributing $1$.

\end{itemize}

This article is organized as follows. In section $2$ we review the construction of noncommutative fuzzy gauge theory on the fuzzy sphere from random matrix theory, and write down our generalized three matrix model. In section $3$ we show by means of Monte Carlo that the emergent fuzzy sphere in this three matrix model is stable for all values of the mass deformation parameter $M$. In section $4$ we write down the analogous six matrix model, and present its one-loop quantization, and then discuss in great detail the calculation of the phase diagram by means of Monte Carlo. It is shown here that the fuzzy four-sphere ${\bf S}^2_N\times{\bf S}^2_N$ is only stable for large values of the mass deformation parameter $M$. In section $5$ we give a detailed discussion of the phases of the model using the eigenvalue distribution. We also revisit in this section the effective potential in order to prove the critical behavior of the theory. In section $6$ various related topics are discussed briefly such as: emergent gauge theory in two dimensions, monopoles and instantons, topology change and stability of the fuzzy four-sphere ${\bf S}^2_N\times{\bf S}^2_N$, critical behavior,  Dirac operators, random matrix theory formulation, and generalization to fuzzy four sphere ${\bf S}^4$ and fuzzy ${\bf CP}^n$. Section $7$ contains our conclusion.

\section{The $3-$dimensional mass deformed Yang-Mills matrix model}
A $U(n)$ gauge action on the fuzzy sphere can be derived from a simple $1-$matrix
model as follows \cite{Steinacker:2003sd}. We introduce Pauli matrices ${\tau}_a$ and the three $N\times N$  matrices $L_a$, which are the $SU(2)$ generators in the irreducible representation of spin $s=(N-1)/{2}$, and define the matrix
\begin{eqnarray}
\bar{C}=(\frac{1}{2}+{\tau}_aL_a){\otimes}{\bf 1}_n.
\end{eqnarray}
It is a trivial exrecise to check that
\begin{eqnarray}
\bar{C}=(j(j+1)-(\frac{N}{2})^2){\otimes}{\bf 1}_n,
\end{eqnarray}
where $j$ is the eigenvalue of the operator $\vec{J}=\vec{L}+{\vec{\sigma}}/{2}$  which takes the two values ${N}/{2}$ and $({N-2})/{2}$. The eigenvalues of $\bar{C}$ are therefore ${N}/{2}$ with multiplicity $n(N+1)$ and $-{N}/{2}$ with multiplicity $n(N-1)$ . Hence
\begin{eqnarray}
\bar{C}^2=(\frac{N}{2})^2{\bf 1}_{2nN}.
\end{eqnarray}
As it turns out this matrix $\bar{C}$ can be obtained as a classical configuration of the following $2nN-$dimensional $1-$matrix action
\begin{eqnarray}
S[C]=\frac{1}{4g^2}\frac{1}{N}Tr tr_2\bigg(C^2-\bigg(\frac{N}{2}\bigg)^2\bigg)^2.\label{ba}
\end{eqnarray}
Indeed, the equations of motion derived from this action reads
\begin{eqnarray}
C(C^2-\frac{N^2}{4})=0.
\end{eqnarray}
It is easy to see that $\bar{C}$ solves this equation of motion and
that the value of the action in this configuration  is identically
zero ,i.e. $S[C=\bar{C}]=0$.

Expanding around the vacuum
$\bar{C}$ by writing 
\begin{eqnarray}
C=\frac{1}{2}+C_0 +{\sigma}_aC_a,
\end{eqnarray}
where $C_0$ and $C_a$ are $nN{\times}nN$ matrices and imposing the
condition 
\begin{eqnarray}
C_0=0,
\end{eqnarray}
we get
\begin{eqnarray}
C^2|_{C_0=0}=\frac{1}{4}+C_a^2+\frac{1}{2}{\epsilon}_{abc}{\sigma}_cF_{ab}.
\end{eqnarray}
The curvature $F_{ab}$ is given in terms of $C_a$ by
\begin{eqnarray}
&&F_{ab}=i[C_a,C_b]+{\epsilon}_{abc}C_c.
\end{eqnarray}
Hence, we obtain the action
\begin{eqnarray}
S[C]=\frac{1}{g^2}\frac{1}{N} \bigg[\frac{1}{4}TrF_{ab}^2+\frac{1}{2}Tr(C_a^2-\frac{N^2-1}{4})^2\bigg].
\end{eqnarray}
The normal scalar field $\Phi$ on the fuzzy sphere is given in terms of $C_a$ by
\begin{eqnarray}
\Phi=\frac{C_a^2-L_a^2}{2\sqrt{c_2}}~,~c_2=L_a^2=\frac{N^2-1}{4}.
\end{eqnarray}
The $U(n)$ gauge action becomes
%Equivalently
%\begin{eqnarray}
%S[C]=\frac{1}{g^2}\frac{1}{N}\bigg[\frac{1}{4}TrF_{ab}^2+2c_2Tr{\Phi}^2\bigg].\label{action9}
%\end{eqnarray}

\begin{eqnarray}
S[C]=\frac{1}{g^2}\frac{1}{N}\bigg[\frac{1}{4}TrF_{ab}^2+2m^2Tr{\Phi}^2\bigg].\label{action9}
\end{eqnarray}
The mass $m^2$ is given by the value of the Casimir, viz
\begin{eqnarray}
m^2=c_2.
\end{eqnarray}
We can bring this
action into the form (with $X_a=2\alpha C_a/3$)
\begin{eqnarray}
&&S[X]
=NTr\bigg[-\frac{1}{4}[X_a,X_b]^2+\frac{2i\alpha}{3}{\epsilon}_{abc}X_aX_bX_c+M(X_a^2)^2+\beta X_a^2\bigg]\nonumber\\
&& M=\frac{m^2}{2c_2}~,~\beta=-\alpha^2\mu=\frac{2{\alpha}^2}{9}(1-2m^2).\label{action00}
\end{eqnarray}
We will also use extensively the parameter $\tilde{\alpha}$ defined by 
\begin{eqnarray}
\tilde{\alpha}^4={\alpha}^4N^2=\bigg(\frac{3}{2}\bigg)^4\frac{1}{g^2}.
\end{eqnarray}
The parameter $g$ is the gauge coupling constant. The classical
absolute minimum of the model is given by the fuzzy sphere
configurations $C_a=L_a$. Expanding around this solution by writing $C_a=L_a+A_a$ yields a $U(n)$ theory  with a $3-$component gauge field $\vec{A}$ where the extra normal component is given by $\Phi$. 
In the commutative limit $N{\longrightarrow}\infty$ the field $\Phi=n_aA_a$ is infinitely heavy and hence it decouples.   The curvature in terms of $A_a$ is given by $F_{ab}=i[L_a,A_b]-i[L_b,A_a]+{\epsilon}_{abc}A_c+i[A_a,A_b]\longrightarrow i{\cal L}_aA_b-i{\cal L}_bA_a+{\epsilon}_{abc}A_c$. The pure gauge action $S[C]$  becomes 
\begin{eqnarray}
S[C]&=&\frac{1}{4g^2N}Tr F_{ab}^2+\frac{2c_2}{g^2N}Tr \Phi^2\longrightarrow\frac{1}{4g^2}\int \frac{d\Omega}{4\pi} F_{ab}^2+\frac{N^2}{2g^2}\int \frac{d\Omega}{4\pi} \Phi^2.
\end{eqnarray}
This is the action of a pure $U(n)$ gauge theory on the ordinary
  sphere. The action (\ref{action00})
  defines therefore  a pure $U(n)$ gauge theory on the fuzzy sphere
   at least in  the region of the phase space where the vacuum configuration
  $C_a=L_a$ is stable. In this study we will deal mostly with  $n=1$ and hence $Tr {\bf
  1}=N$.

The action (\ref{action00}) should be compared with the action $(3.8)$ of \cite{DelgadilloBlando:2008vi}. Here $m^2=c_2$ whereas  $m^2$ is a free parameter in \cite{DelgadilloBlando:2008vi}. Furthermore, $\beta$ is fixed here as $\beta=2\alpha^2/9$ for $m^2=0$  whereas $\beta$  is a free parameter for $m^2=0$ in \cite{DelgadilloBlando:2008vi}. This action is also slightly different from the action $(1)$ studied in \cite{O'Connor:2006wv}. Here, the $m^2=0$ theory is the perfect square action $F_{ab}^2/2$ whereas the $m^2=0$ limit of the action $(1)$ studied in \cite{O'Connor:2006wv} is the Alekseev-Recknagel-Schomerus (ARS) stringy action given by $F_{ab}^2/2+$ Chern-Simons or equivalently \cite{Alekseev:2000fd}
\begin{eqnarray}
&&S[X]
=NTr\bigg[-\frac{1}{4}[X_a,X_b]^2+\frac{2i\alpha}{3}{\epsilon}_{abc}X_aX_bX_c\bigg].
\end{eqnarray}
The perfect square action $F_{ab}^2/2$ and the ARS action are connected by the critical line in the plane $\tilde{\alpha}-\beta$  of the model with $M=0$ computed in \cite{DelgadilloBlando:2012xg}.

\section{An emergent stable fuzzy sphere}
We want to study, by means of Monte Carlo, the phase diagram in the plane $\tilde{\alpha}-M$  of the following model

\begin{eqnarray}
&&S[X]
=NTr\bigg[-\frac{1}{4}[X_a,X_b]^2+\frac{2i\alpha}{3}{\epsilon}_{abc}X_aX_bX_c+M(X_a^2)^2+\beta X_a^2\bigg].\label{fuzzys2}
\end{eqnarray}
The parameter $M$ is taken in the range between $M=0$ (perfect square action) and $M=1/2$ (Steinacker's action) and even beyond whereas $\beta$ is given in terms of $M$ by 
\begin{eqnarray}
\beta=-\alpha^2\mu=\frac{2{\alpha}^2}{9}(1-4c_2M).
\end{eqnarray}
The background minimal solution  of this model  is $X_a=\alpha\phi L_a$. In the limit $m^2\longrightarrow \infty$, we have $\mu\longrightarrow m^2$, and we find a critical line separating the fuzzy sphere phase solution with $\phi\neq 0$, from the Yang-Mills matrix phase solution with $\phi=0$, given by the critical line 
%The theoretical prediction is given by 
\begin{eqnarray}
\tilde{\alpha}^4=\frac{81}{2m^2}.%\Rightarrow \bar{\alpha}=\frac{2}{M^{1/4}}.
\end{eqnarray}
We use the Metropolis algorithm for the Monte Carlo update.

\begin{figure}[htbp]
\begin{center}
\includegraphics[width=13.0cm, angle=0]{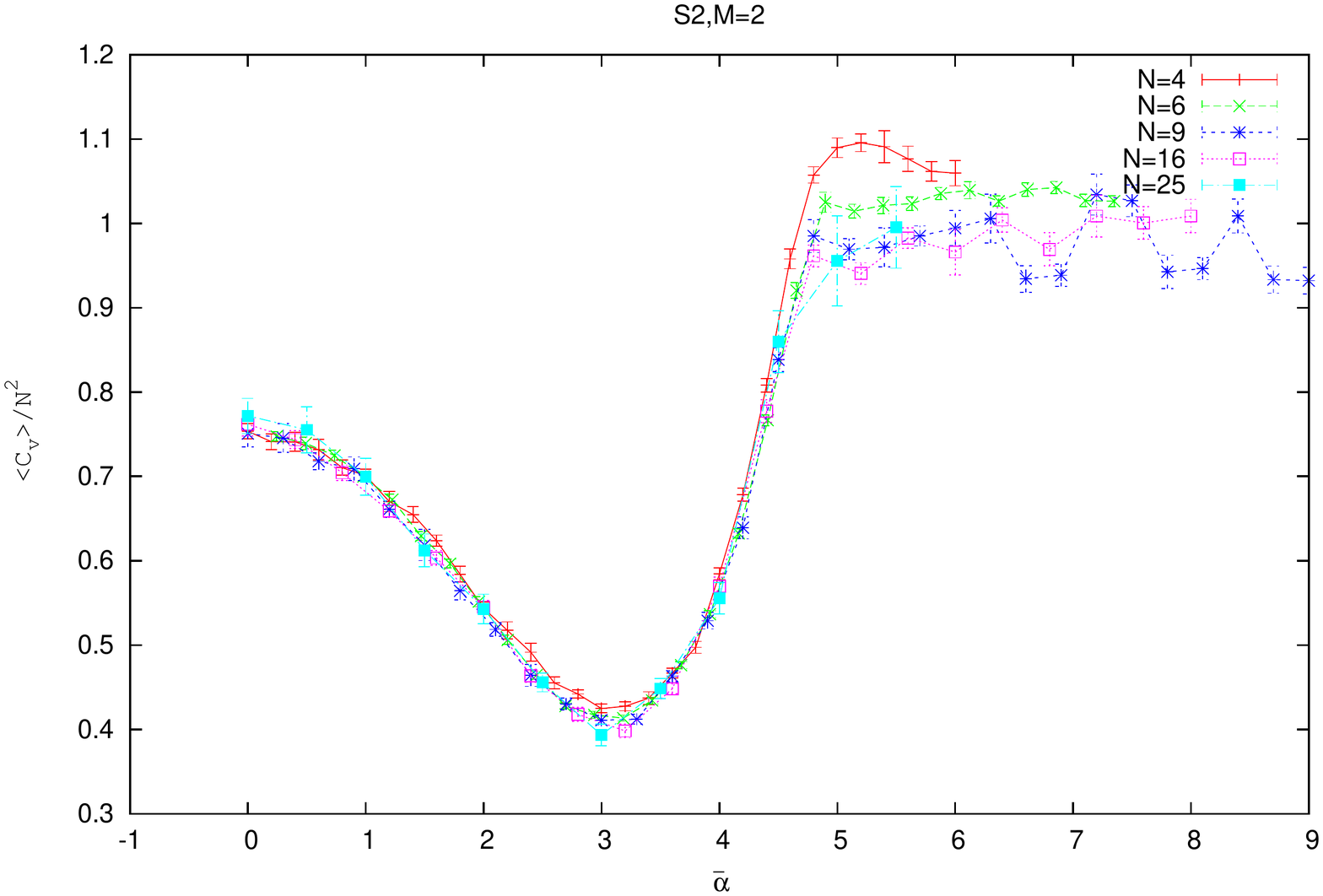}\\
\includegraphics[width=13.0cm,angle=0]{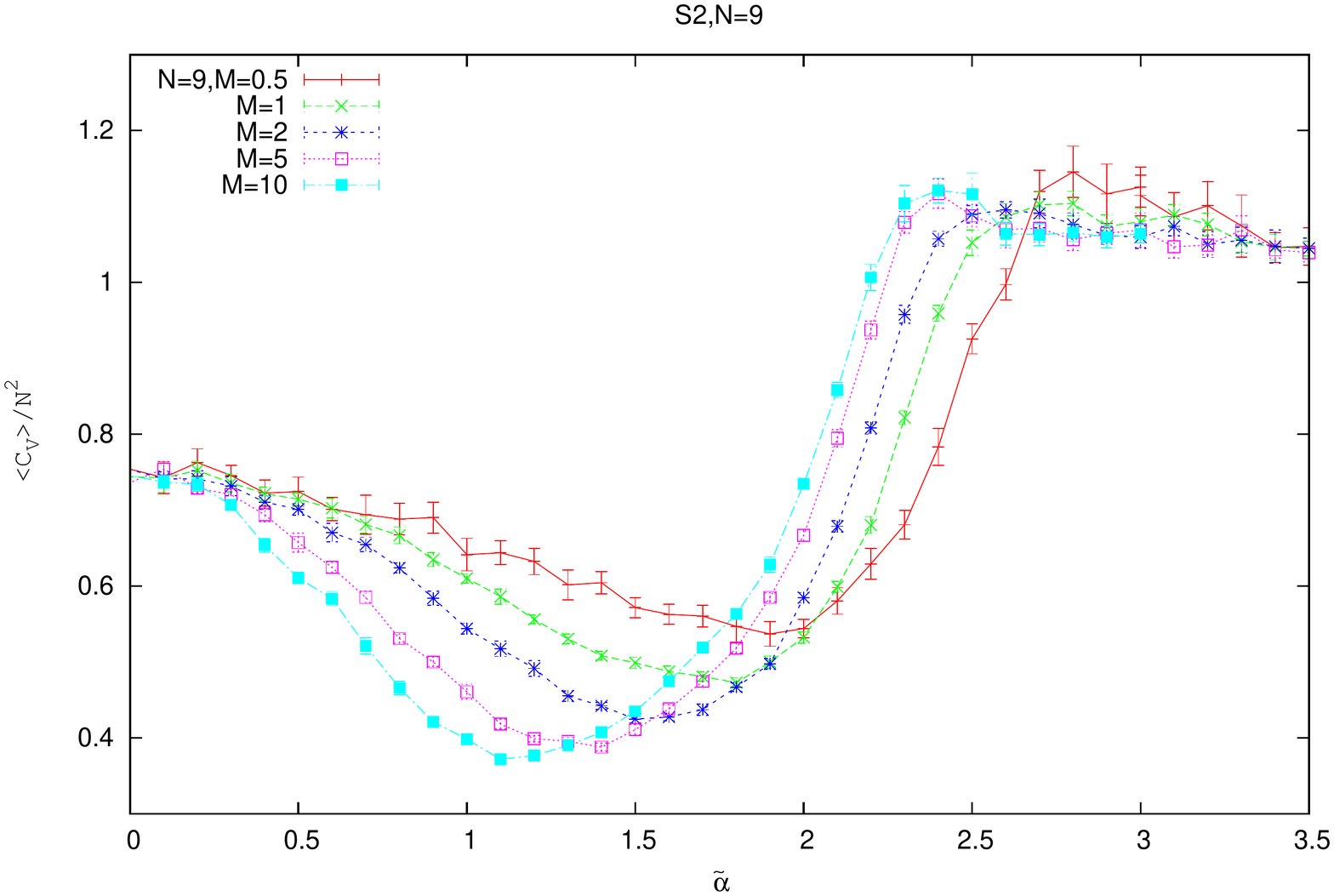}
\end{center}
\caption{The specific heat in the three dimensional Yang-Mills matrix model.}\label{cv}
\end{figure}

\begin{figure}[htbp]
\begin{center}
\includegraphics[width=15.0cm,angle=0]{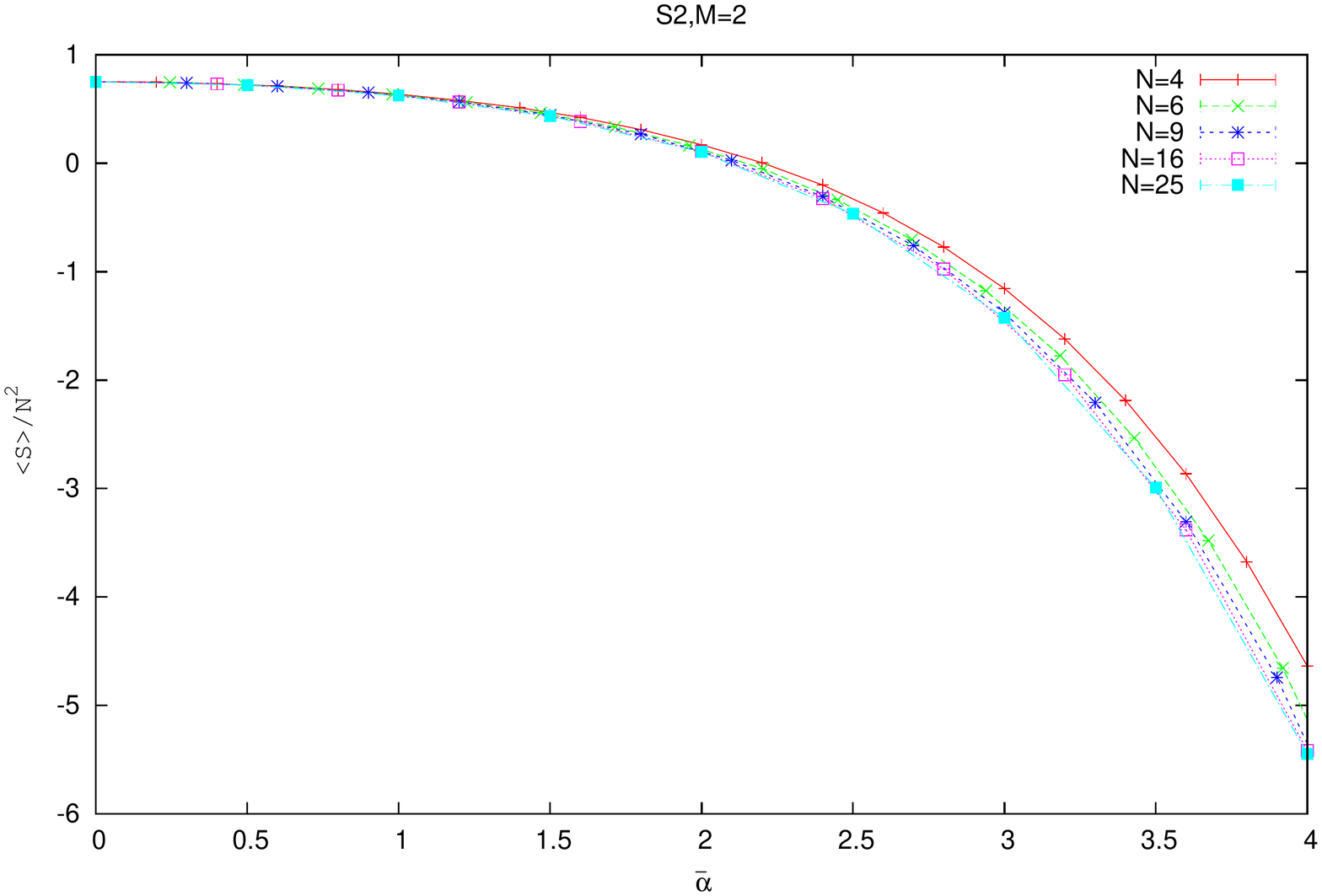}\\
\includegraphics[width=15.0cm,angle=0]{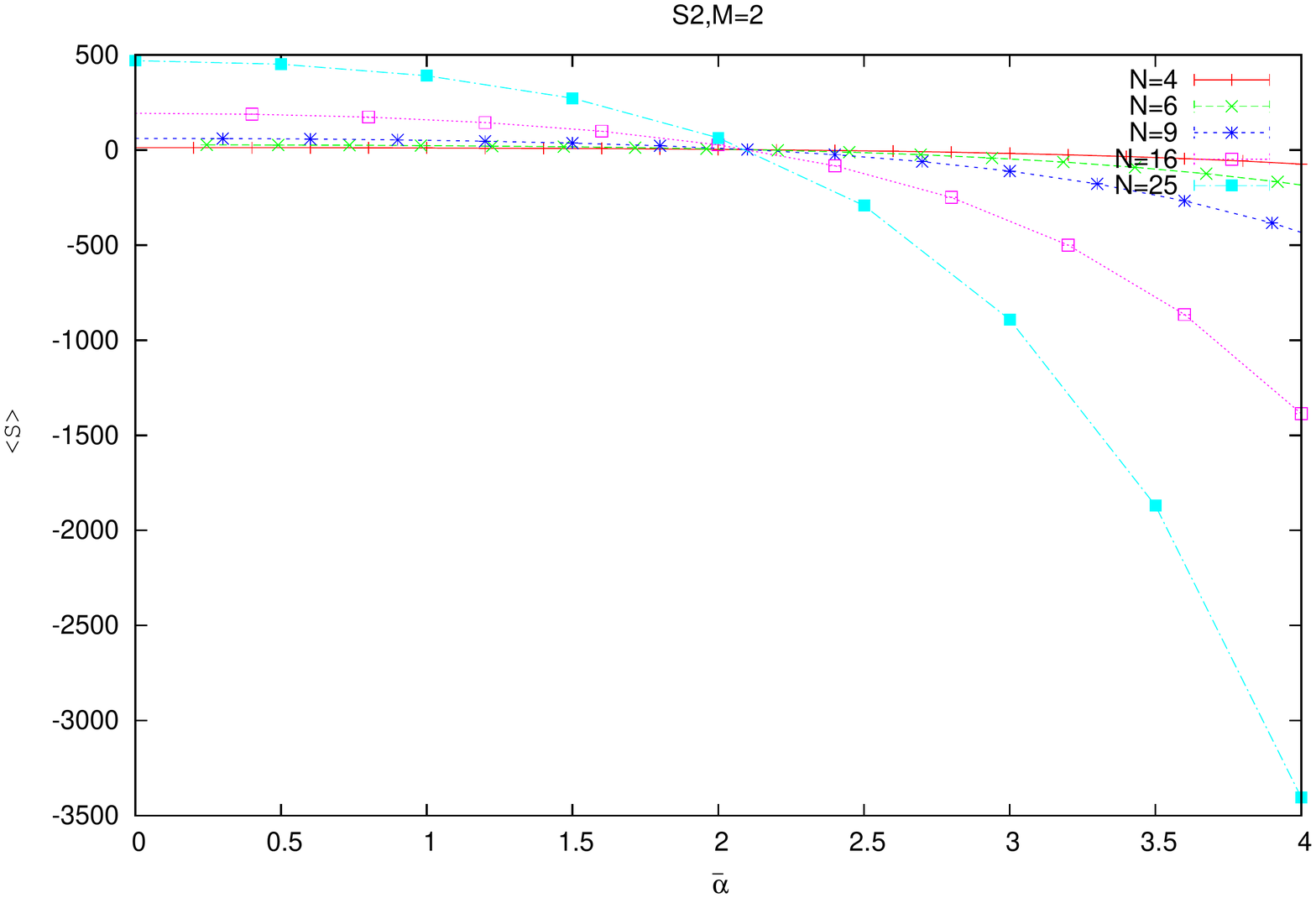}
\end{center}
\caption{The action in the three dimensional Yang-Mills matrix model.}\label{action}
\end{figure}

For each value of the mass parameter $M$, we have measured the critical point $\tilde{\alpha}_*$ at the minimum of the specific heat 
\begin{eqnarray}
C_v=<S^2>-<S>^2.
\end{eqnarray}
with error bars estimated by the value of the step. See figure (\ref{cv}). The results are shown on table (\ref{1}). We observe from table (\ref{2}) that the collapsed coupling constant is not $\tilde{\alpha}$ but it is given by
\begin{eqnarray}
\bar{\alpha}=\tilde{\alpha}\sqrt{N}=\alpha N.
\end{eqnarray}
Another measurement of the critical value $\bar{\alpha}_*$ is given by intersection point of the actions with different values of $N$. See figure (\ref{action}). The results are shown on table (\ref{3}). 

\begin{figure}[htbp]
\begin{center}
\includegraphics[width=13.0cm, angle=0]{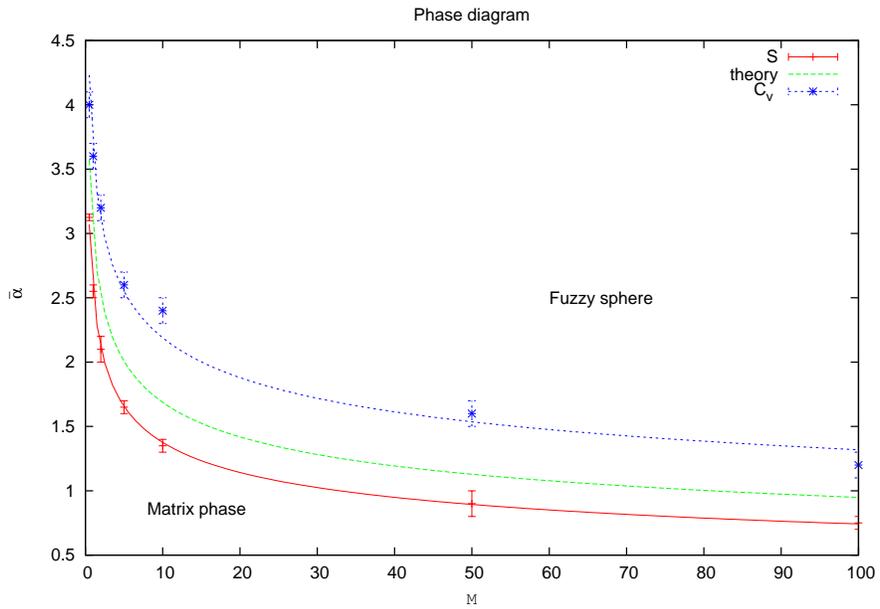}\\
\includegraphics[width=13.0cm, angle=0]{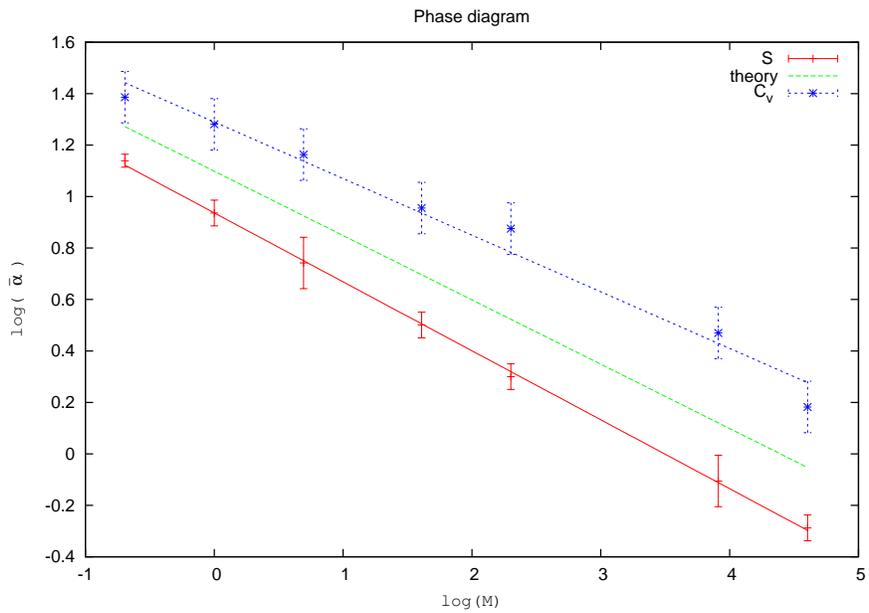}
\end{center}
\caption{The phase diagram of the three dimensional Yang-Mills matrix model.}\label{phasediagram}
\end{figure}

The two different measurements, which give an upper and lower estimates, of the critical point $\bar{\alpha}_*$ are included on the phase diagram (\ref{phasediagram}). The data are fitted to straight lines and compared to theory in table (\ref{fit}).  The theoretical prediction in terms of the collapsed coupling constant $\bar{\alpha}$ is given by 
\begin{eqnarray}
\bar{\alpha}=\frac{3}{M^{1/4}}.
\end{eqnarray}
This shows explicitly that the fuzzy sphere is absolutely stable in this model for all values of $M$, including the Steinacker value $M=1/2$, since the inverse of the critical gauge coupling constant behaves as 
\begin{eqnarray}
\tilde{\alpha}=\frac{3}{N^{1/2}M^{1/4}}\longrightarrow 0.\label{stable_sphere}
\end{eqnarray}
The sphere-to-matrix transition line is pushed to $0$ and only one phase survives. This is the result advocated originally by Steinacker in \cite{Steinacker:2003sd}.

\begin{figure}[htbp]
\begin{center}
\includegraphics[width=15cm,angle=0]{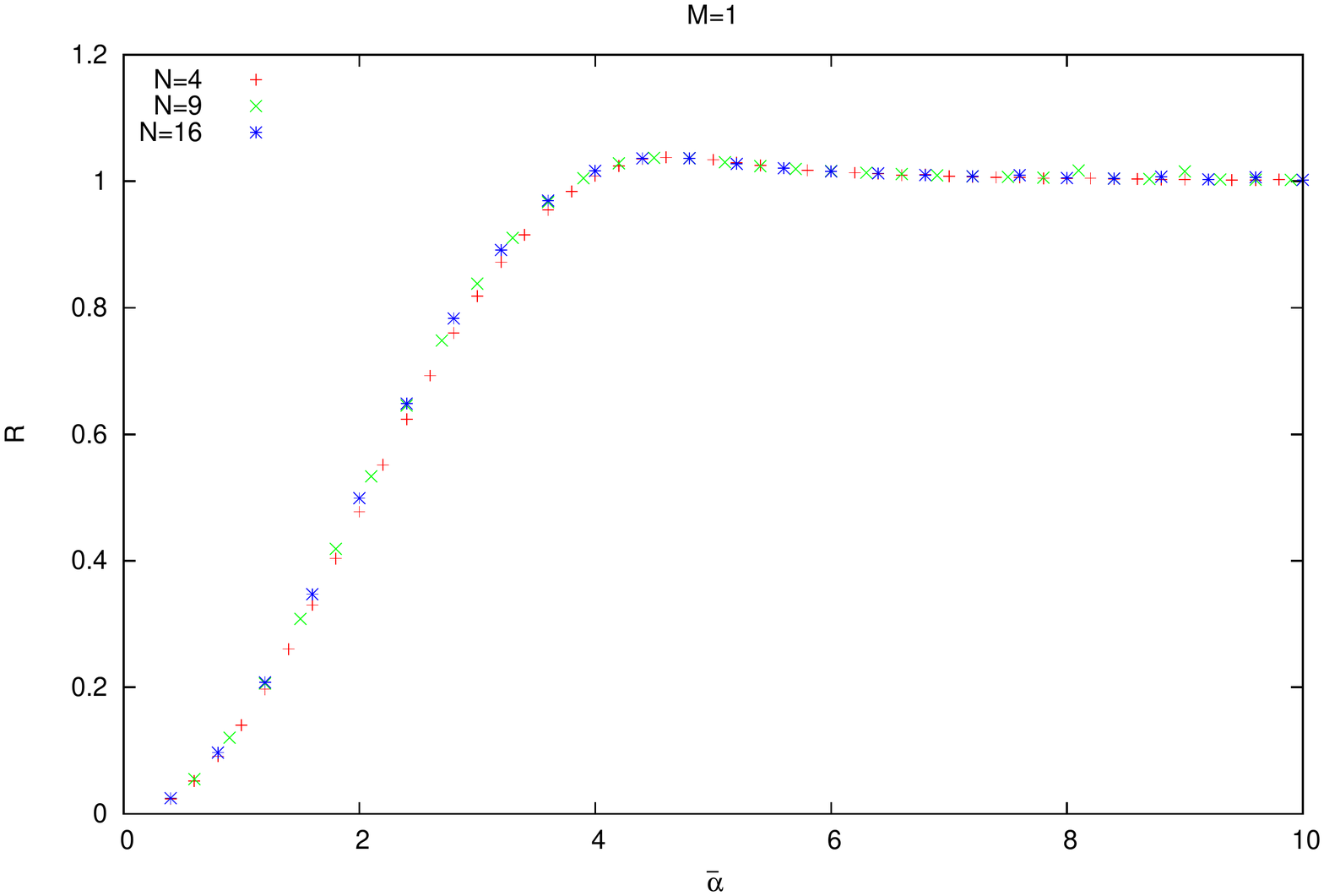}\\
\includegraphics[width=15cm,angle=0]{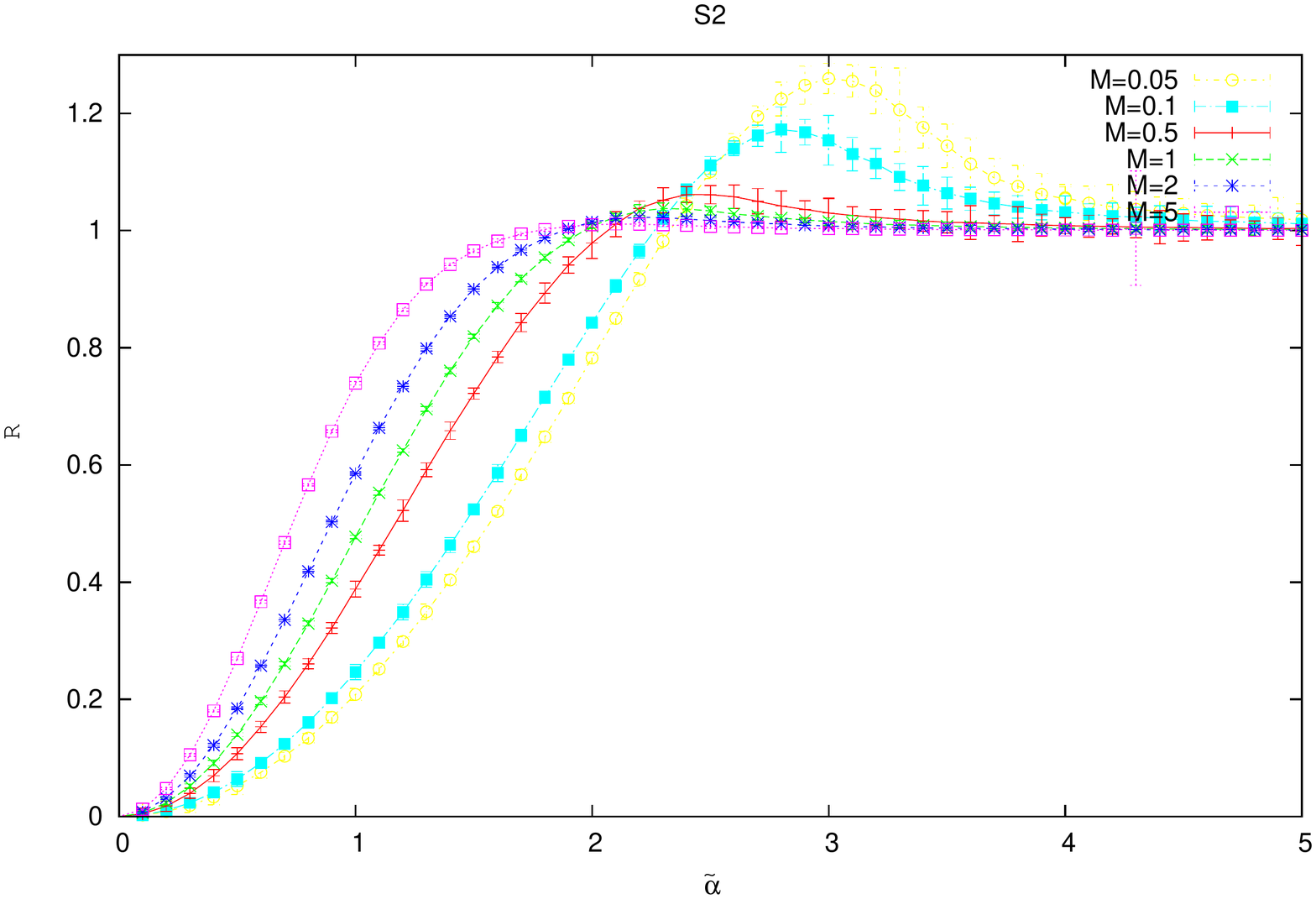}
\end{center}
\caption{The radius in the three dimensional Yang-Mills matrix model.}\label{radius}
\end{figure}

For completeness we also measure the radius of the sphere $R$ as a function of the mass $M$ and $N$. This is defined by the formula  
\begin{eqnarray}
&&\frac{1}{R}=\frac{1}{\phi^2 \tilde{\alpha}^2 c_2} Tr X_a^2~,~c_2=\frac{N^2-1}{4} ,\phi=\frac{2}{3}.\label{radiuss}
\end{eqnarray}
The results are shown on figure (\ref{radius}).

\begin{table}
\begin{center}
%\centering
\begin{tabular}{ |c|c|c|c|c|c| } 
 \hline
 $M/\tilde{\alpha}$ & $N=4$ &$N=6$ & $N=9$ &$N=16$  &$ N=25$\\  
 \hline
 $ 0.5$ &$1.9\pm 0.1$&$1.5\pm 0.1$  &$1.3 \pm 0.1 $ &$ 1.0\pm 0.1  $ & $0.8\pm 0.1 $\\ 
\hline
 $1 $ &$  1.7\pm 0.1 $&$1.45\pm 0.05$  & $  1.2\pm 0.1 $ &$ 0.9\pm 0.1  $ &$0.7\pm 0.1 $ \\
 \hline
$2$ &$  1.6\pm 0.1 $&$1.4\pm 0.1 $ & $ 1.0 \pm 0.1  $ &$ 0.8\pm 0.1$ & $ 0.6\pm 0.1$ \\
 \hline
$5$ &$ 1.3 \pm 0.1 $&$1.1\pm 0.1 $ & $ 0.9\pm 0.1  $ &$ 0.65\pm 0.05$  &$0.55\pm0.1 $  \\
 \hline
$10$ &$  1.2\pm 0.1 $&$1\pm 0.1$ & $ 0.8 \pm 0.1 $ &$0.6\pm 0.1 $ & $0.5\pm 0.1 $  \\
 \hline
$50$ &$  0.8\pm 0.1 $&$0.7\pm 0.1$ & $0.5\pm 0.1 $&$0.4\pm 0.1 $ & $0.3\pm 0.1 $  \\
 \hline
$100$ &$  0.7\pm 0.1 $&$0.5\pm 0.1$ & $0.4\pm 0.1 $ &$0.3\pm 0.1 $ & $0.3\pm 0.1 $  \\
 \hline
\end{tabular}
\end{center}
\caption{The critical values $\tilde{\alpha}_*$ for different values of $M$ and $N$.}\label{1}
\end{table}

\begin{table}
\begin{center}
\begin{tabular}{ |c|c|c|c|c|c| } 
 \hline
 $M/\bar{\alpha}$ & $N=4$ &$N=6$ & $N=9$ &$N=16$  &$ N=25$\\ 
 \hline
 $ 0.5$ &$3.8\pm 0.1$&$3.7\pm 0.1$  &$3.9 \pm 0.1 $ &$ 4.0\pm 0.1 $ & $4.0\pm 0.1  $ \\ 
\hline
 $1 $ &$  3.4\pm 0.1 $&$3.55\pm 0.05$  & $  3.6\pm 0.1 $ &$ 3.6\pm 0.1  $ &$3.5\pm0.1 $  \\
 \hline
$2$ &$  3.2\pm 0.1 $&$3.4\pm 0.1 $ & $ 3 \pm 0.1  $ &$ 3.2\pm 0.1$ & $ 3\pm 0.1$ \\
 \hline
$5$ &$ 2.6 \pm 0.1 $&$2.7\pm 0.1 $ & $ 2.7\pm 0.1  $ &$ 2.6\pm 0.05$  &$2.75\pm 0.05 $  \\
 \hline
$10$ &$  2.4\pm 0.1 $&$2.4\pm 0.1$ & $ 2.4 \pm 0.1 $ &$2.4\pm 0.1 $ & $2.5\pm 0.1 $  \\
 \hline
$50$ &$1.6\pm 0.1$&$1.7\pm 0.1$ & $1.5\pm0.1 $ &$1.6\pm 0.1 $ & $1.5\pm 0.1 $  \\
 \hline
$100$ &$1.4\pm0.1$&$1.22\pm 0.1$ & $ 1.2\pm 0.1 $ &$1.2\pm 0.1 $ & $1.5\pm 0.1 $  \\
 \hline
\end{tabular}
\end{center}
\caption{The critical values $\bar{\alpha}_*$ for different values of $M$ and $N$.}\label{2}
\end{table}

\begin{table}
\begin{center}
\begin{tabular}{ |c|c| } 
 \hline
 $M$&$\bar{\alpha}$  \\ 
 \hline
 $ 0.5$ &$ 3.125$\\
\hline
 $1 $ & $2.55$\\
 \hline
 $2 $ &$2.1 $\\
 \hline
 $5 $ &$1.65 $\\
 \hline
 $10 $ &$1.35   $\\
 \hline
 $50 $ &$0.9   $\\
 \hline
 $100 $ &$0.75  $\\
 \hline
\end{tabular}
\end{center}
\caption{The critical values $\bar{\alpha}_*$ from the intersection point of the actions.}\label{3}
\end{table}

\begin{table}
\begin{center}
\begin{tabular}{ |c|c|c| } 
 \hline
 method &$ a $ & $b$  \\ 
 \hline
Minimum $ C_v $ &$ -0.22 \pm0.014 $&$1.29\pm0.036$\\
\hline
 Theory & $-0.25  $& $0.48 $\\
 \hline
Intersection $S $ &$-0.268 \pm0.003 $&$0.936  \pm 0.007$\\
 \hline
\end{tabular}
\end{center}
\caption{The slop and the intercept of the straight lines used to fit the data with the theoretical prediction.}\label{fit}
\end{table}
%\begin{figure}[htbp]
%\begin{center}
%\includegraphics[angle=0,scale=0.45]{data_rouag/cvsymm.pdf}
%\end{center}
%\caption{$C_v$}
%\end{figure}

\section{The emergent fuzzy four-sphere ${\bf S}^2\times {\bf S}^2$ is stable in the limit $M\longrightarrow\infty$}
\subsection{The $6-$dimensional mass deformed Yang-Mills matrix model}
Generalization of the above construction is straightforward to four dimensions, i.e. to fuzzy ${\bf S}^2\times{\bf S}^2$, and yields immediately the $6-$matrix action
 \begin{eqnarray}
S[X,Y]
&=&NTr\bigg[-\frac{1}{4}[X_a,X_b]^2+\frac{2i\alpha}{3}{\epsilon}_{abc}X_aX_bX_c+M(X_a^2)^2+\beta X_a^2\bigg]\nonumber\\
&+&NTr\bigg[-\frac{1}{4}[Y_a,Y_b]^2+\frac{2i\alpha}{3}{\epsilon}_{abc}Y_aY_bY_c+M(Y_a^2)^2+\beta Y_a^2\bigg]\nonumber\\
&+&NTr\bigg[-\frac{1}{2}[X_a,Y_b]^2\bigg].\label{action00}
\end{eqnarray}
We choose $N$ to be a perfect square such that 
\begin{eqnarray}
N=N_0^2.
\end{eqnarray}
The parameters of the model are given by 
\begin{eqnarray}
M=\frac{m^2}{2c_2^{0}}~,~\beta=-\alpha^2\mu~,~\mu=\frac{2}{9}(4c_{2}^0M-1)~,~c_2^0=\frac{N_0^2-1}{4}.
\end{eqnarray}
Before we report the Monte Carlo data we discuss the one-loop effective potential of this theory and its phase structure. 

The background solution (absolute minimum) of this model is by construction given by
\begin{eqnarray}
X_a=\alpha\phi L_a\otimes {\bf 1}~,~Y_a=\alpha\phi {\bf 1}\otimes L_a.
\end{eqnarray}
The fuzzy four-sphere is given explicitly by 
\begin{eqnarray}
x_1^2+x_2^2+x_3^2=1~,~
[x_a,x_b]=\frac{i}{\sqrt{c_2^0}}{\epsilon}_{abc}x_c,
\end{eqnarray}
\begin{eqnarray}
y_1^2+y_2^2+y_3^2=1~,~
[y_a,y_b]=\frac{i}{\sqrt{c_2^0}}{\epsilon}_{abc}y_c,
\end{eqnarray}
\begin{eqnarray}
[x_a,y_b]=0,
\end{eqnarray}
where $x_a=L_a\otimes {\bf 1}/\sqrt{c_2^0}$ and $y_a={\bf 1}\otimes L_a/\sqrt{c_2^0}$.

The $L_a$ are the $SU(2)$ generators in the irreducible representation of spin $s=(N_0-1)/2$, and the background value of $\phi$ is $2/3$. By expanding around this solution, as $X_a=2\alpha C_a^{(1)}/3$, $Y_a=2\alpha C_a^{(2)}/3$, $C_a^{(i)}=L_a^{(i)}+A_a^{(i)}$, we obtain in the limit $N\longrightarrow\infty$ a $U(1)$ gauge theory on ${\bf S}^2\times{\bf S}^2$ given by the action 
\begin{eqnarray}
S[C^{(1)},C^{(2)}]=\frac{1}{g^2}\frac{1}{N}Tr \bigg[\frac{1}{4}F_{ab}^{(1)2}+\frac{1}{4}F_{ab}^{(2)2}+\frac{1}{2}F_{ab}^{(12)2}+2m^2{\Phi}^{(1)2}+2m^2{\Phi}^{(2)2}\bigg].
\end{eqnarray}
By comparing with equation $(20)$ of \cite{Behr:2005wp}  it seems to us that $m^2=c_2$ as in the case of the single fuzzy sphere. The definition of the components of curvature tensor and the two scalar fields are obviously given by

\begin{eqnarray}
&&F_{ab}^{(i)}=i[C_a^{(i)},C_b^{(i)}]+{\epsilon}_{abc}C_c^{(i)}~,~F_{ab}^{(12)}=i[C_a^{(1)},C_b^{(2)}].
\end{eqnarray}
\begin{eqnarray}
\Phi^{(i)}=\frac{C_a^{(i)2}-c_2^0}{2\sqrt{c_2^0}}.
\end{eqnarray}
\subsection{Quantization at one-loop}
For simplicity let us go back to the case of a single sphere and discuss the derivation of the effective potential there first \cite{CastroVillarreal:2004vh}. 

Quantization around this background, using the background field method gives, after gauge fixing in the Lorentz gauge, the effective action
 \begin{eqnarray}
{\Gamma}[X_a]&=&S[X_a]+\frac{1}{2}Tr\log{\Omega}-Tr\log{\cal X}^2,\label{gamma}
\end{eqnarray}
where  the Laplacian operator ${\Omega}$ is given explicitly by the formula 
\begin{eqnarray}
{\Omega}_{ab}&=& {\cal X
}_c^2{\delta}_{ab}-2{\cal
F}_{ab}+4M (X_c^2-c_2\alpha^2){\delta}_{ab}+8M D_aD_b+2(\beta+2c_2\alpha^2M) {\delta}_{ab}.
\end{eqnarray}
The first term in (\ref{gamma}) is due to the gauge field while the second term is due to the ghost field. In above the notation ${\cal X}_a$ and ${\cal F}_{ab}$ means that
the covariant derivative $X_a$ and the curvature $F_{ab}=i[X_a,X_b]+\alpha\epsilon_{abc}X_c$ act by
commutators, i.e ${\cal X}_a(A)=[X_a,A]$, ${\cal
F}_{ab}(A)=[F_{ab},A]$ where $A{\in}Mat_{N}$. Similarly,
${\cal X}^2(A)=[X_a,[X_a,A]]$. 

The UV-IR mixing behavior on the fuzzy four-sphere in the model with $M=\beta=0$ was studied in great detail in \cite{DelgadilloBlando:2006dp}. 

The effective potential for $X_a=\alpha\phi L_a$, by neglecting the contributions from all the terms in the Laplacian $\Omega$ except the first, is given by
\begin{eqnarray}
\frac{V}{2c_2}=\tilde{\alpha}^4\bigg[\frac{\phi^4}{4}-\frac{{\phi}^3}{3}+m^2\frac{\phi^4}{4}-\mu\frac{\phi^2}{2}\bigg]+\log\phi^2.
\end{eqnarray}
The calculation on the fuzzy four-sphere ${\bf S}^2\times{\bf S}^2$ proceeds in exactly the same way \cite{CastroVillarreal:2005uu,DelgadilloBlando:2006dp}. First we fix the Lorentz gauge, then compute the effective action, and then substitute the background matrices $X_a=\alpha\phi L_a\otimes {\bf 1}$, $Y_a=\alpha\phi {\bf 1}\otimes L_a$. The calculation of the classical part of the effective potential is trivial. The quantum part goes along the same lines as above. In particular, the contributions from all the terms in the Laplacian $\Omega$ are negligible except the first one. Thus we get the logarithmic potential 
\begin{eqnarray}
\frac{1}{2}Tr_dTr\log\phi^2-Tr\phi^2=\frac{d}{2}N^2\log\phi^2-N^2\log\phi^2.
\end{eqnarray}
On ${\bf S}^2$ we have $d=3$ whereas on ${\bf S}^2\times {\bf S}^2$ we have $d=6$. We get then on  ${\bf S}^2\times{\bf S}^2$ the effective  potential 
\begin{eqnarray}
\frac{V}{2N^2}&=&2c_{2}^{0}{\alpha}^4\bigg[\frac{\phi^4}{4}-\frac{{\phi}^3}{3}+m^2\frac{\phi^4}{4}-\mu\frac{\phi^2}{2}\bigg]+\log\phi^2\nonumber\\
&=&\tilde{\alpha}_0^4\bigg[\frac{\phi^4}{4}-\frac{{\phi}^3}{3}+m^2\frac{\phi^4}{4}-\mu\frac{\phi^2}{2}\bigg]+\log\phi^2,
\end{eqnarray}
where we have redefined the coupling constant by 
 \begin{eqnarray}
\frac{N_0^2}{2}{\alpha}^4=\tilde{\alpha}_0^4.
\end{eqnarray}
The difference between the result on ${\bf S}^2$ and this result lies in the replacement $\tilde{\alpha}\longrightarrow\tilde{\alpha}_0$ and the replacement $c_2\longrightarrow c_2^0$ in the definition of $\mu$. The analysis of the phase structure is therefore identical (see section $5.2$). 

The equation of motion reads then
\begin{eqnarray}
\frac{V^{'}}{2N^2}
&=&\tilde{\alpha}_0^4\bigg[\phi^3-{\phi}^2+m^2\phi^3-\mu \phi\bigg]+\frac{2}{\phi}.
\end{eqnarray}
In the limit $m^2\longrightarrow \infty$, we have $\mu\longrightarrow 4m^2/9$, and we find a critical line separating the fuzzy sphere phase solution with $\phi\neq 0$, from the Yang-Mills matrix phase solution with $\phi=0$, given by the formula
\begin{eqnarray}
&&(1+m^2){\phi}_*=\frac{3}{8}\big(1+\sqrt{1+\frac{32t}{9}}\big)\nonumber\\
&&\frac{1}{\tilde{\alpha}_{0*}^4}=\frac{\phi_*^2(\phi_*+2\mu)}{8}\nonumber\\
&&t=\mu(1+m^2).
\end{eqnarray}
More detail on the derivation of this formula and the analysis of the phase diagram from the effective potential can be found in section $5.2$. 

The small mass limit $M\longrightarrow 0$ gives 
\begin{eqnarray}
\tilde{\alpha}\sim N
\end{eqnarray}
which diverges with $N$. This is precisely the divergent asymptotic behavior of the critical line of the model $M=0$ observed in \cite{DelgadilloBlando:2012xg}.
 
The large mass limit $M\longrightarrow\infty$ (or equivalently $m^2\longrightarrow \infty$) of these equations is given in terms of $\tilde{\alpha}$ by

\begin{eqnarray}
\tilde{\alpha}=3\big(\frac{2}{M}\big)^{1/4}.\label{theo}
\end{eqnarray}
Thus as opposed to the case of a single sphere the collapsed coupling constant is $\tilde{\alpha}$ and not $\bar{\alpha}$ which signals the persistence of the instability of the emergent four-sphere geometry as we will now show with the Monte Carlo results. 
\subsection{Monte Carlo calculation of the phase diagram}
We perform Monte Carlo simulation of the above six dimensional matrix model using the Metropolis algorithm. We vary $M$ for different values of $N$. We work with $N=4-25$ and $M=0.05-50$. 

\begin{figure}[htbp]
\begin{center}
\includegraphics[width=14.0cm,angle=0]{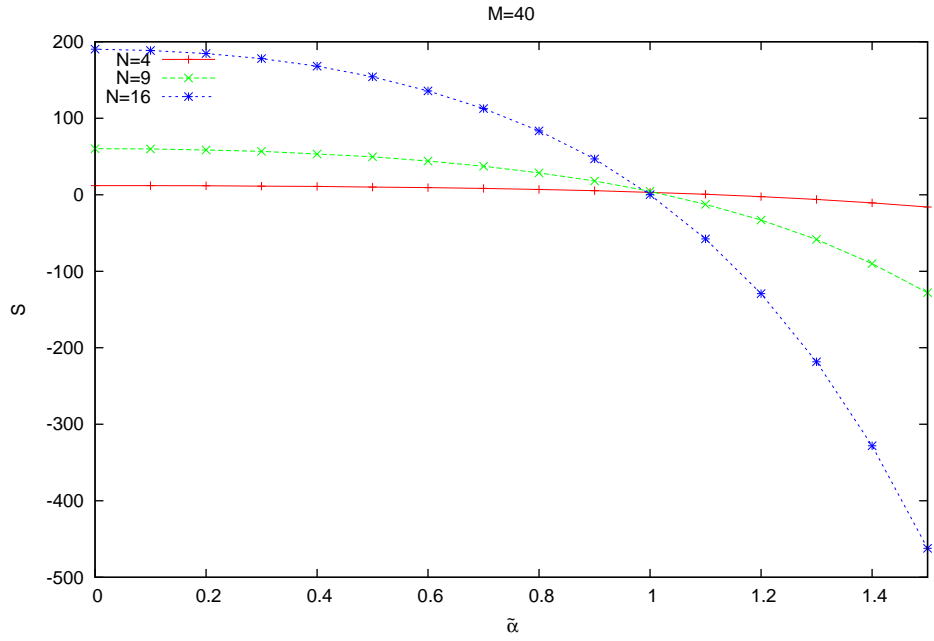}\\
\includegraphics[width=14.0cm,angle=0]{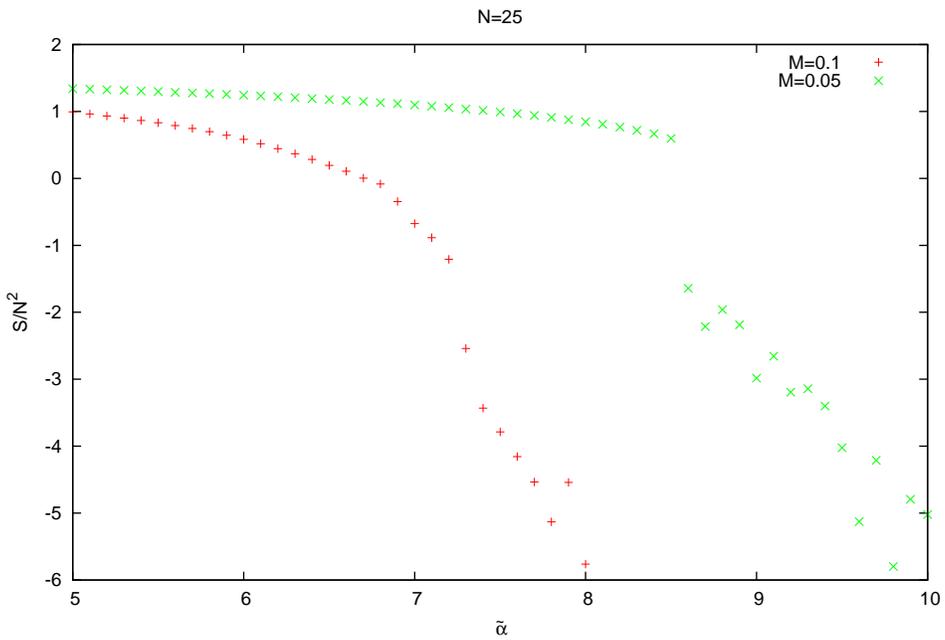}\\
\caption{The intersection point of the action in the six dimensional Yang-Mills matrix model.}\label{actions2s2}
\end{center}
\end{figure}

\begin{figure}[htbp]
\begin{center}
\includegraphics[width=14cm,angle=0]{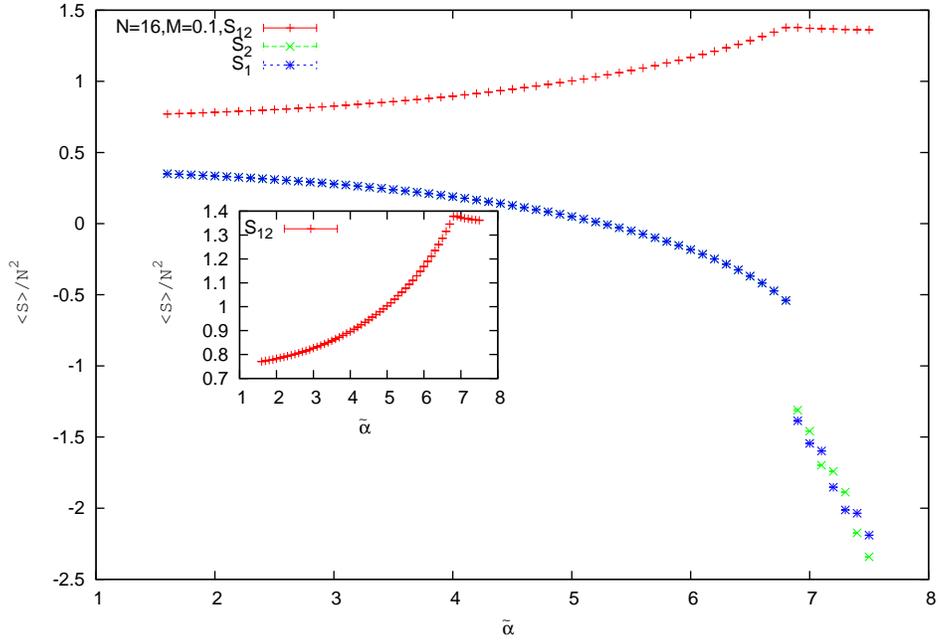}
\includegraphics[width=14cm,angle=0]{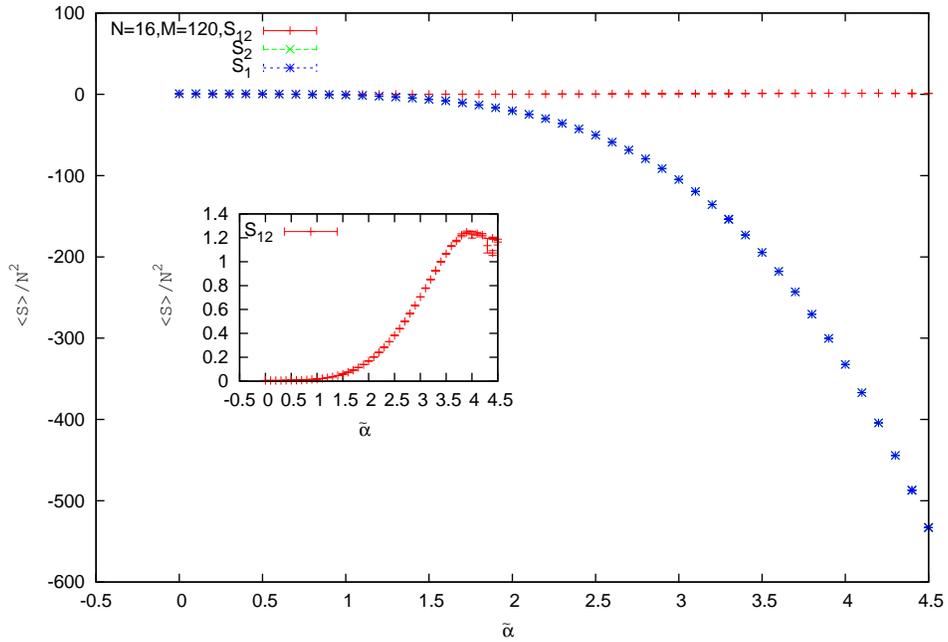}
\end{center}
\caption{The action ${\rm YM}_{12}$.}\label{YM12}
\end{figure}

\begin{figure}[htbp]
\begin{center}
\includegraphics[width=14.0cm,angle=0]{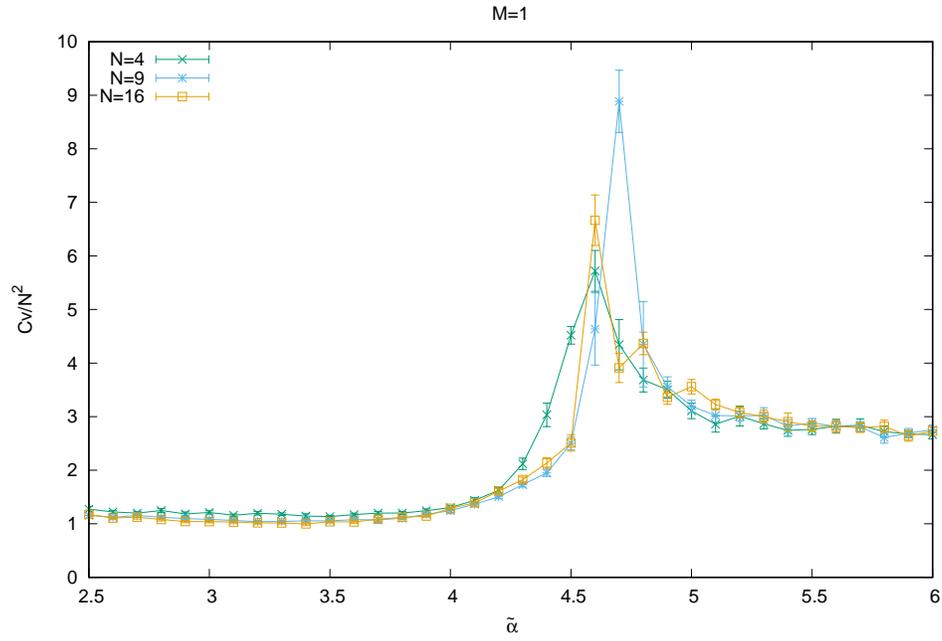}\\
\includegraphics[width=14.0cm,angle=0]{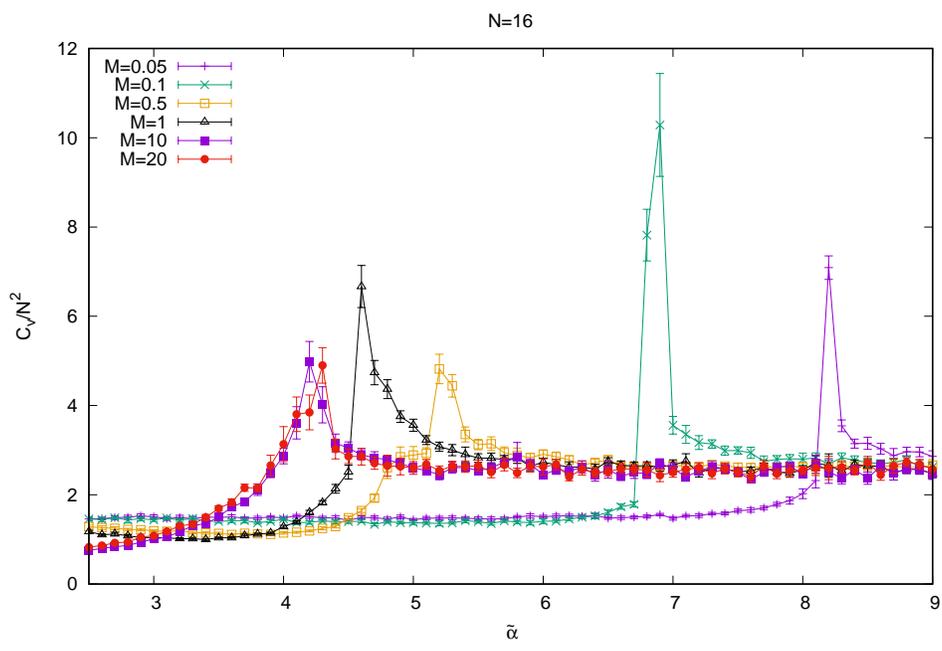}
\caption{The specific heat (peak) in the six dimensional Yang-Mills matrix model.}\label{cvs2s2}
\end{center}
\end{figure}

\begin{figure}[htbp]
\begin{center}
\includegraphics[width=14.0cm,angle=0]{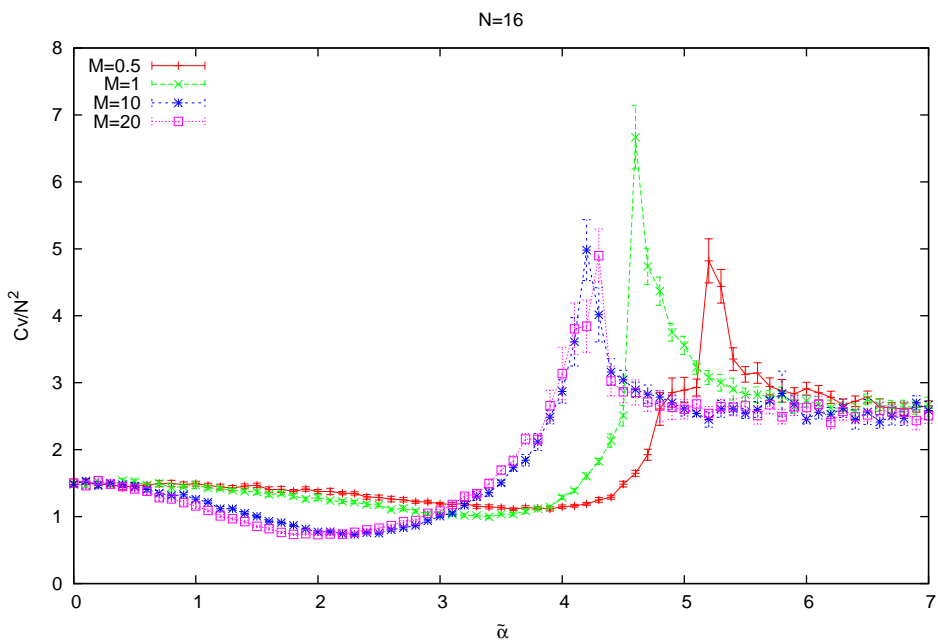}
\caption{The specific heat (minimum) in the six dimensional Yang-Mills matrix model.}\label{cvs2s2e}
\end{center}
\end{figure}

\begin{figure}[h]
\begin{center}
\includegraphics[width=14cm,angle=0]{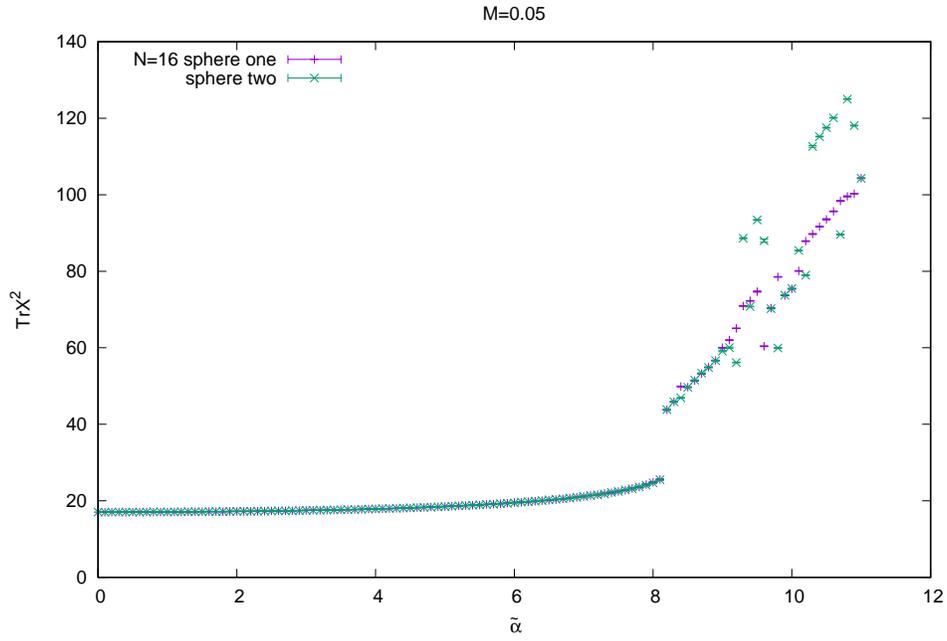}\\
\includegraphics[width=14cm,angle=0]{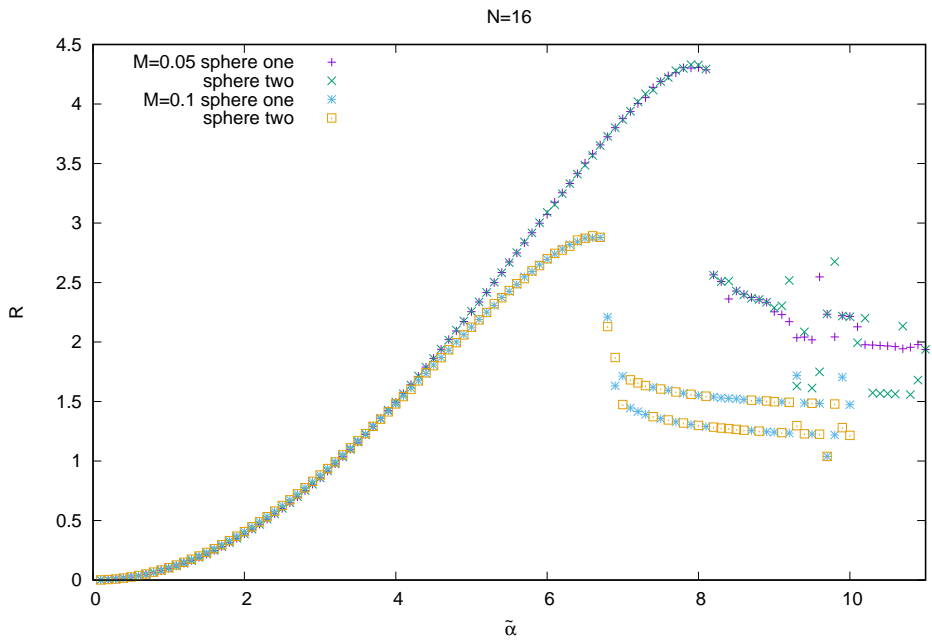} 
\caption{The radius in the six dimensional Yang-Mills matrix model for small values of $M$.}\label{radiuss2s23}
\end{center}
\end{figure}

\paragraph{The action:}
\begin{itemize}
\item The intersection point of the average actions (entropies) for various values of $N$ leads in this case to a robust measurement of the transition point between the fuzzy four-sphere ${\bf S}^2\times{\bf S}^2$ phase and the matrix Yang-Mills phase. See (\ref{actions2s2}). This measurement in comparison with the theoretical prediction value (\ref{theo}) gives an under estimation of the critical point. The results are included on table (\ref{s2s2table2e}).

\item For small values of $M$ we also observe a discrete jump in the entropy at the transition point between the fuzzy four-sphere ${\bf S}^2\times{\bf S}^2$ phase and the matrix Yang-Mills phase. See the second graph of(\ref{actions2s2}). The physics of this discontinuity is discussed in more detail in section $5$ using the approximation of the effective potential. 

\item The behavior of the action which couples the two spheres, viz 
\begin{eqnarray}
S_{12}={\rm YM}_{12}=-\frac{N}{2}Tr[X_a,Y_b]^2,
\end{eqnarray}
for small and large values of $M$, is shown on figure (\ref{YM12}). We choose $M=0.1$, and $M=120$, for $N=16$. We also plot $S_1$ and $S_2$, which are obviously defined, for comparison.

We observe that for small values of $M$ the action ${\rm YM}_{12}$ is comparable to $S_i$, whereas for large values of $M$ it becomes quite negligible. This means in particular, that for large values of $M$, the approximation of the effective potential is expected to work well.
\end{itemize}
\paragraph{The specific heat:}
\begin{itemize}
\item For small values of $M$, the geometric phase transition from the fuzzy four-sphere ${\bf S}^2\times{\bf S}^2$ phase to the matrix Yang-Mills phase is marked by, and is measured at, the peak of the specific heat. See figure (\ref{cvs2s2}). The peak becomes harder to resolve for larger values of $M$ and $N$. This is in contrast with the case of the fuzzy two-sphere ${\bf S}^2$ where a peak in the specific heat is observed only for very small values of $M$. The results are included on table  (\ref{s2s2table1}).

\item There seems to exist a new transition in $M$, around $M\sim 0.5$ which is the Steinacker's value, beyond which the peak in $C_v$ ceases from marking the transition from the fuzzy four-sphere phase to the Yang-Mills matrix phase, and the specific heat develops a minimum where the transition actually occurs. See figure (\ref{cvs2s2e}) and the results are included on table  (\ref{s2s2table1e}). Thus, for larger values of $M$ above $M\sim 0.5$ the  geometric phase transition from the fuzzy four-sphere ${\bf S}^2\times{\bf S}^2$ phase to the matrix Yang-Mills phase is marked by, and is measured at, the minimum of $C_v$. 

\item We also observe that the value at the peak of the specific heat saturates at around $\tilde{\alpha}\sim 4.1$ for larger values of $M$. This peak corresponds in this case to a transition from a fuzzy four-sphere to a crossover phase as we will describe further shortly.
\end{itemize}
\paragraph{The radius:}
We also measure the radii of the two spheres given in terms of $TrX_a^2$ and $TrY_a^2$ respectively by the formula (\ref{radiuss}) with the substitution $N\longrightarrow N_0$. From the Monte Carlo data we observe no difference, beyond and above statistical fluctuations, between the two radii. This is obvious from the figure (\ref{radiuss2s23}).

We observe for small values of $M$ a discontinuity in the radius at the transition point as seen neatly on figure (\ref{radiuss2s23}). We note that in this case it becomes harder to thermalize the system in the fuzzy four-sphere phase due to the zero modes of the matrices $X_a$ and $Y_a$.

\begin{figure}[h]
\begin{center}
\includegraphics[width=14cm,angle=0]{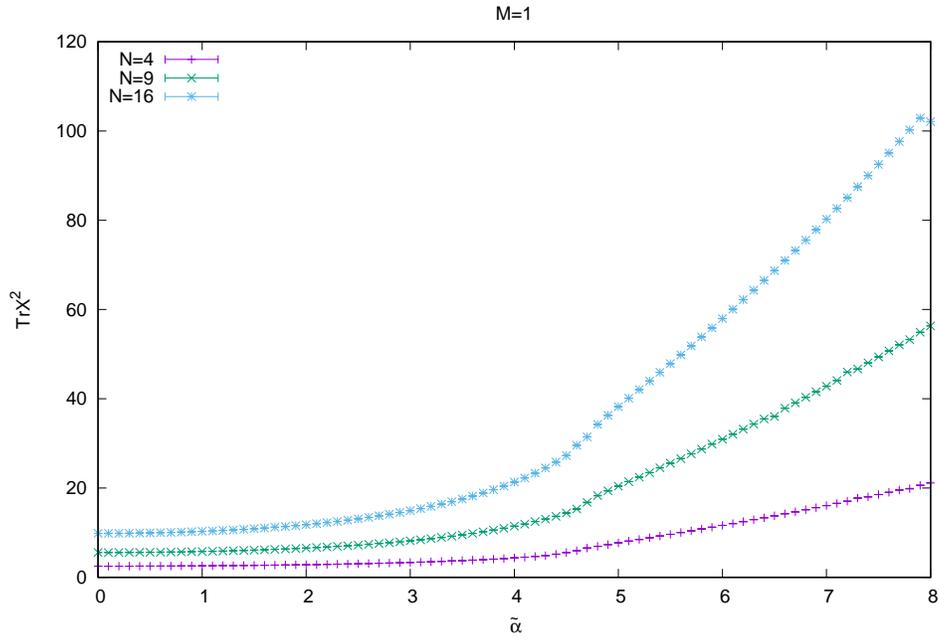}\\
\includegraphics[width=14cm,angle=0]{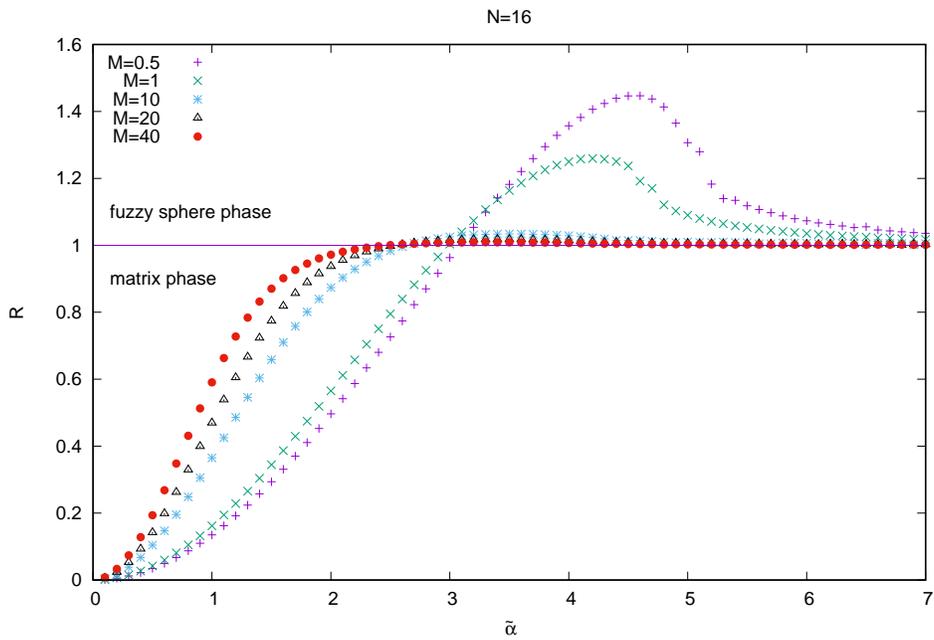}
\caption{The radius in the six dimensional Yang-Mills matrix model.}\label{radiuss2s21}
\end{center}
\end{figure}

For medium and larger values of $M$ the discontinuity is smoothed out as shown on figure (\ref{radiuss2s21}). The transition point in this case is taken, for medium values of $M$, at the maximum reached by the radius $R$ in the fuzzy four-sphere phase before decreasing to zero in the Yang-mills matrix phase.  For larger values of $M$, the transition point  is taken at the point where the radius $R$ drops below one. In summary,

\begin{equation*}
    R  \longrightarrow \begin{cases}
                
                1~,~\tilde{\alpha}>>\tilde{\alpha}_*               & \text{fuzzy four-sphere phase}\\
               0~,~\tilde{\alpha}<<\tilde{\alpha}_*                & \text{Yang-Mills matrix phase.}\\
%                \text{except the small M, the point of transition is the max of R .}
           \end{cases}
\end{equation*}
This consists an independent measurement of the critical point between the fuzzy four-sphere phase and the Yang-Mills matrix phase. The data for different values of $N$ is well collapsed and thus one can obtain from the radius a single estimate for the critical point shown on table (\ref{s2s2table2}).

\begin{figure}[h]
\begin{center}
 \includegraphics[width=15.0cm,height=9.0cm,angle=0]{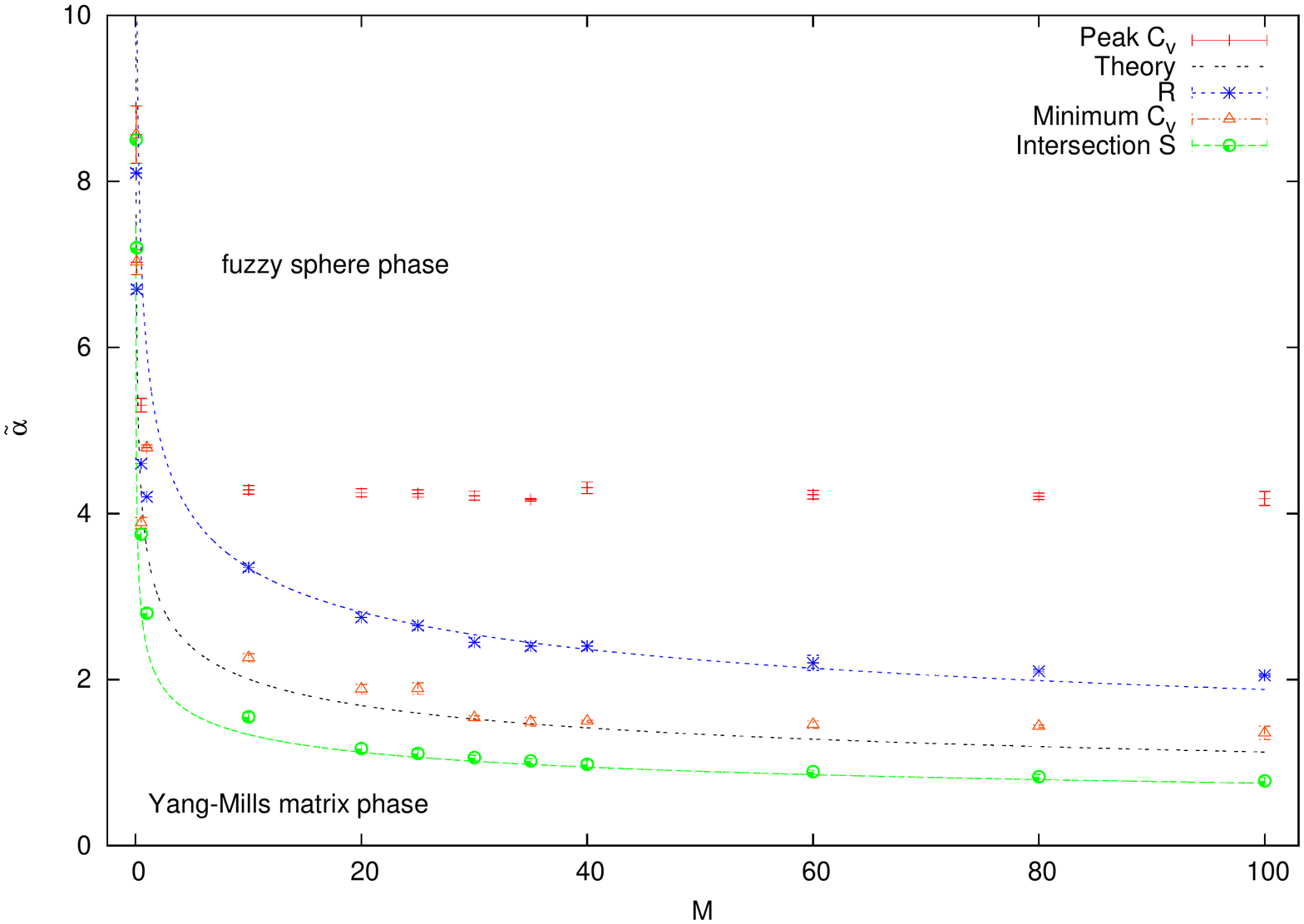}
 \includegraphics[width=15.0cm,height=9.0cm,angle=0]{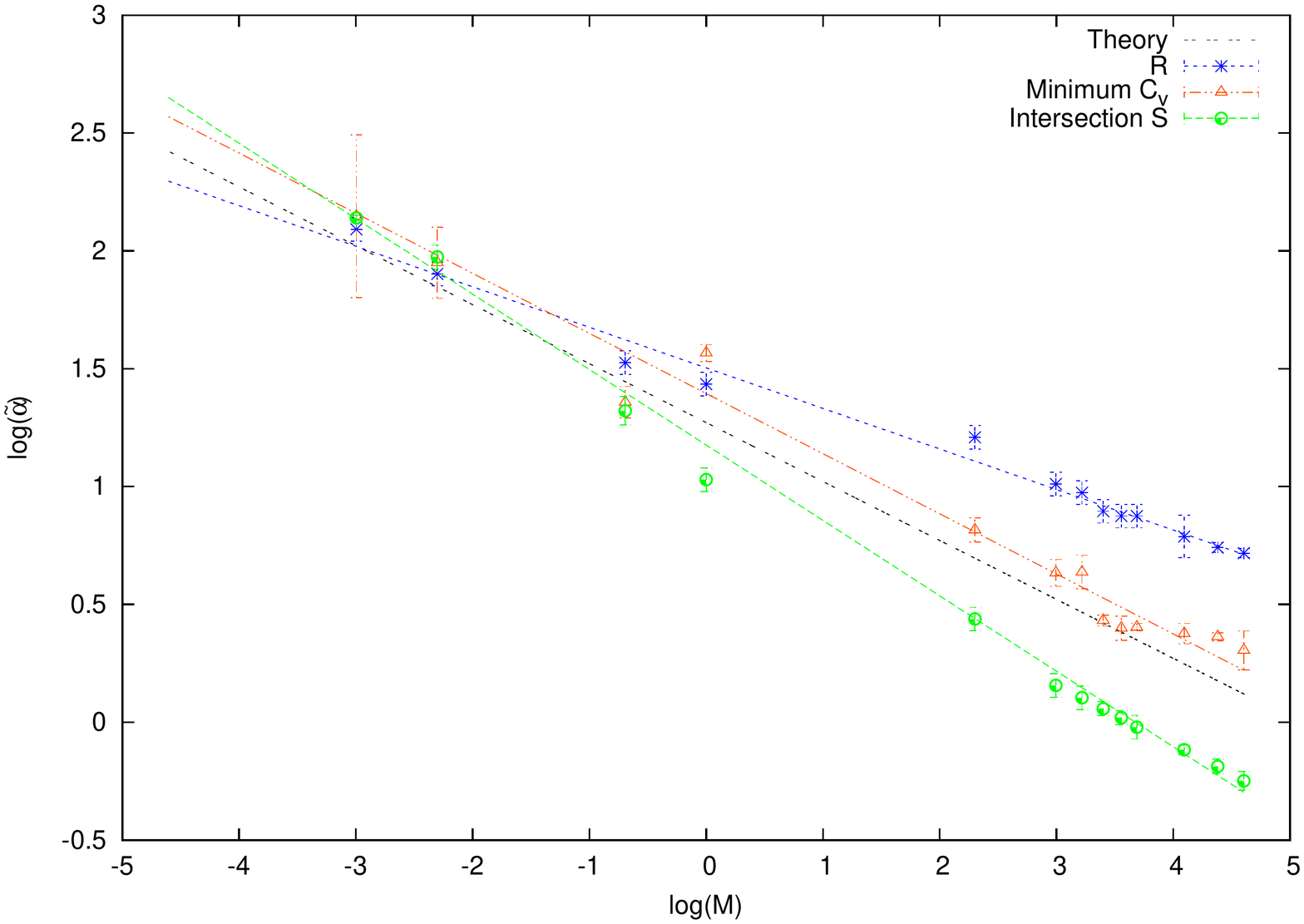}
\caption{The phase diagram of the six dimensional Yang-Mills matrix model.}\label{phasediagram2}
\end{center}
\end{figure}

\paragraph{The phase diagram:}
The critical line, and as a consequence the phase diagram, from the measurements of the action (intersection point or jump), the specific heat (peak and minimum) and the radius (jump, maximum and dropping below $1$), together with the theoretical calculation given by equation (\ref{theo}),  are shown on figure  (\ref{phasediagram2}).

In summary, we observe roughly three phases. The fuzzy four-sphere phase as expected for large values of $\tilde{\alpha}$, and a Yang-Mills matrix phase for small values of $\tilde{\alpha}$, and this is the case for every value of the mass parameter $M$. 

Therefore, in this case the fuzzy four-sphere phase is only stable in the limit $M\longrightarrow \infty$ since the collapsed gauge coupling constant is  $\tilde{\alpha}$ and not $\bar{\alpha}$, in contrast to what happens in the case of a single sphere. 

As we will discuss, in the next section, the Yang-Mills phase in this case is characterized by different eigenvalue distributions for large and small values of $M$. There seems to exist another transition in $M$ around the Steinacker's value $M=0.5$ where the profile of the eigenvalue distribution, for $\tilde{\alpha}=0$, changes from the $d=6$ law (small values of $M$) to a uniform distribution (large value of $M$). See below for a detailed discussion.

Also, we observe that the phase diagram develops a distinct third phase between the   fuzzy four-sphere phase and the Yang-Mills matrix phase for large values of $M$. This looks like a crossover phase. The critical line between the  fuzzy four-sphere phase and this new phase, as measured by the peak of $C_v$, saturates around $\tilde{\alpha}=4.2$. Another possible  interpretation of this phase is that of a strongly coupled gauge theory on the emergent background geometry and this is supported by the measurement of the radius in this region.

\begin{table}
\centering
\begin{tabular}{|l|c|}
\hline
$M$              &  $\tilde{\alpha}$\\

\hline
$M=0.05$ & $8.50$ \\
\hline
$M=0.1$ & $7.20$ \\
\hline
$M=0.5$ & $3.75$ \\
\hline
$M=1.0$ & $2.80$ \\
\hline
$M=10$ & $1.55$ \\
\hline
$M=20$ & $1.17$ \\
\hline
$M=25$ & $1.11$ \\
\hline
$M=30$ & $1.06$ \\
\hline
$M=35$ & $1.02$ \\
\hline
$M=40$ & $0.98$\\
\hline
$M=60$ & $0.89$ \\
\hline
$M=80$ & $0.83$ \\
\hline
$M=100$ & $0.78$\\
\hline
\end{tabular}
\caption{The critical values $\tilde{\alpha}_*$ from the intersection of the action. }
\label{s2s2table2e}
\end{table}

%\label{radiuss2s21}
\begin{table}
\centering
\begin{tabular}{|l|c|c|c|c|c|}
\hline
$ M/\tilde{\alpha}$             & $N=4$  & $N=9$   & $N=16$ & $N=25$ & Extrapolation\\

\hline

$M=0.05$ & $7.50$ &$7.60$  &$8.20$  &$8.70$ &$8.5614 \pm 0.345$\\
\hline
$M=0.1$ & $6.10$ &$6.70$  &$6.80$  &$7.00$ &$7.0262\pm 0.1505  $\\
\hline
$M=0.5$ & $4.90$ & $5.00$ &$5.20$  &$5.30$ &$5.3040 \pm  0.0813$\\
\hline
$M=1.0$ & $4.60$ & $4.70$ &$4.70$  &$4.80$ &$4.7914 \pm 0.0360$\\
\hline
$M=10$ & $4.20$ & $4.30$ &$4.20$  & $4.30$ &$4.2851 \pm 0.0534$\\
\hline
$M=20$ & $4.20$ & $4.20$ & $4.30$ & $4.20$ &$4.2482 \pm 0.0492$\\
\hline
$M=25$ & $4.30$ & $4.30$ & $4.30$ &$4.20$ &$4.2419 \pm 0.0449$\\
\hline
$M=30$ & $4.30$ & $4.20$ & $4.30$  &$4.20$ &$4.2149 \pm 0.0534$\\
\hline
$M=35$ & $4.30$ & $4.20$ & $4.20$  &$4.20$ &$4.1666 \pm 0.0168$\\
\hline
$M=40$ & $4.10$ & $4.30$ & $4.30$  &$4.20$ &$4.3087 \pm 0.0717$\\
\hline
$M=60$ & $4.20$ & $4.30$ & $ 4.20$  &$4.20$ &$4.2271 \pm 0.0531$\\
\hline
$M=80$ & $4.10$ & $ 4.10$ & $4.20$  &$4.20$ &$4.2063 \pm 0.0374$\\
\hline
$M=100$ & $4.20$ & $4.30$ & $4.10$  &$4.20$ &$4.1788 \pm 0.0849$\\
\hline
\end{tabular}
\caption{The critical values $\tilde{\alpha}_*$ from the peak in $C_v$.}
\label{s2s2table1}
\end{table}

%\label{radiuss2s21}
\begin{table}
\centering
\begin{tabular}{|l|c|c|c|c|c|}
\hline
$ M/\tilde{\alpha}$             & $N=4$  & $N=9$   & $N=16$ & $N=25$ & Extrapolation\\

\hline

$M=0.05$ & $7.50$ &$7.60$  &$8.20$  &$8.70$ &$8.5614 \pm 0.345$\\
\hline
$M=0.1$ & $6.10$ &$6.70$  &$6.80$  &$7.00$ &$7.0262\pm 0.1505  $\\
\hline
$M=0.5$ & $3.70$ & $3.70$ &$3.85$  &$3.90$ &$3.8885\pm   0.0664$\\
\hline
$M=1.0$ & $4.60$ & $4.70$ &$4.70$  &$4.80$ &$4.7914 \pm 0.0360$\\
\hline
$M=10$ & $2.50$ & $2.30$ &$2.30$  &$2.35$ &$2.2622 \pm 0.05124$\\
\hline
$M=20$ & $2.20$ & $2.00$ & $1.90$  &$2.00$ &$1.8850 \pm 0.0571$\\
\hline
$M=25$ & $2.10$ & $1.90$ & $1.90$  & $2.00$&$1.8912 \pm 0.0717$\\
\hline
$M=30$ & $2.10$ & $1.80$ & $1.70$  & $1.80$&$1.5404 \pm 0.0233$\\
\hline
$M=35$ & $2.00$ & $1.80$ & $1.65$  &$1.65$ &$1.4917 \pm 0.0516$\\
\hline
$M=40$ & $1.90$ & $1.70$ & $1.60$  & $1.55$&$1.4979 \pm 0.0155$\\
\hline
$M=60$ & $1.70$ & $1.50$ & $1.55 $  & $1.50$&$1.4573 \pm 0.0433$\\
\hline
$M=80$ & $1.60$ & $1.50 $ & $1.50$  & $1.45$&$ 1.4376 \pm 0.0168$\\
\hline
$M=100$ & $1.60$ & $1.40$ & $1.45$  & $1.40$&$1.3573 \pm 0.0823$\\
\hline
\end{tabular}
\caption{The critical values $\tilde{\alpha}_*$ from the minimum in $C_v$.}
\label{s2s2table1e}
\end{table}

\begin{table}
\centering
\begin{tabular}{|l|c|}
\hline
$M$              &  $\tilde{\alpha}$\\

\hline

$M=0.05$ & $8.10$ \\
\hline
$M=0.1$ & $6.70$ \\
\hline
$M=0.5$ & $4.60$ \\
\hline
$M=1.0$ & $4.20$ \\
\hline
$M=10$ & $3.50$ \\
\hline
$M=20$ & $2.50$ \\
\hline
$M=25$ & $2.50$ \\
\hline
$M=30$ & $2.45$ \\
\hline
$M=35$ & $2.40$ \\
\hline
$M=40$ & $2.40 $\\
\hline
\end{tabular}
\caption{The critical values $\tilde{\alpha}_*$ from the radius. }
\label{s2s2table2}
\end{table}

\section{Eigenvalues distributions and critical behavior}
\subsection{Eigenvalues distributions from Monte Carlo}
The phase transition between the fuzzy four-sphere phase and the Yang-Mills phase can be characterized fully by the behavior of the eigenvalue distribution across the transition line. This is by far the most detailed order parameter at our disposal.

\begin{figure}[htbp]
\begin{center}
\includegraphics[width=14.0cm,angle=0]{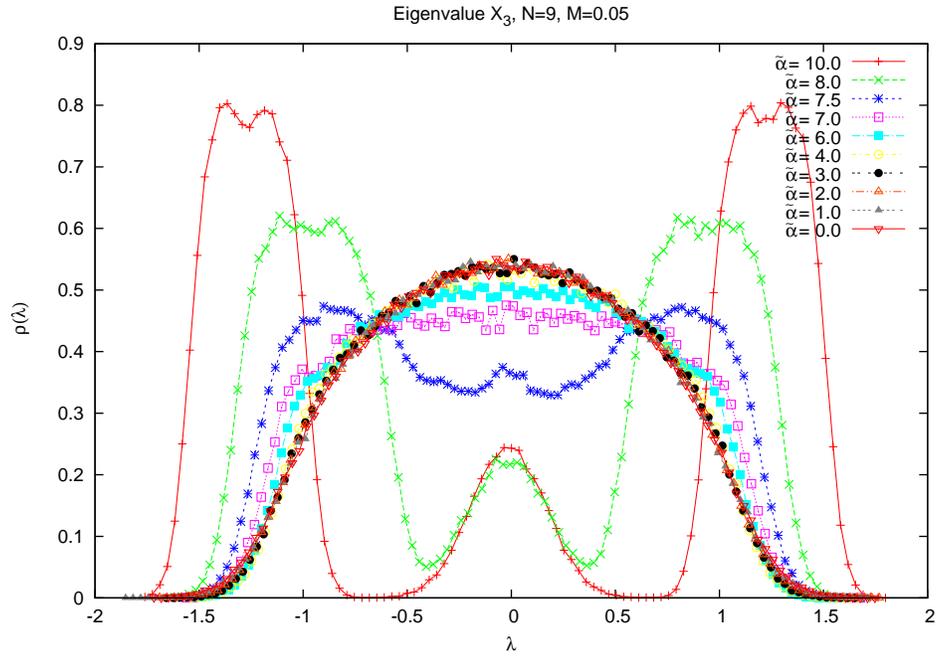}
\includegraphics[width=14.0cm,angle=0]{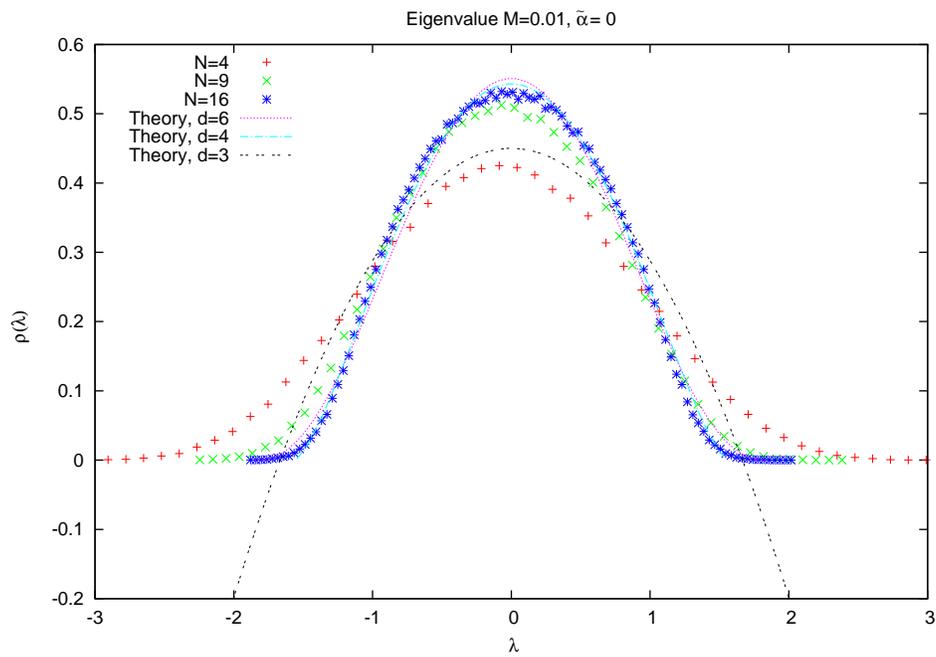}
\caption{The eigenvalue distributions for small values of $M$.}\label{ev1}
\end{center}
\end{figure}

\begin{figure}[htbp]
\begin{center}
\includegraphics[width=14.0cm,angle=0]{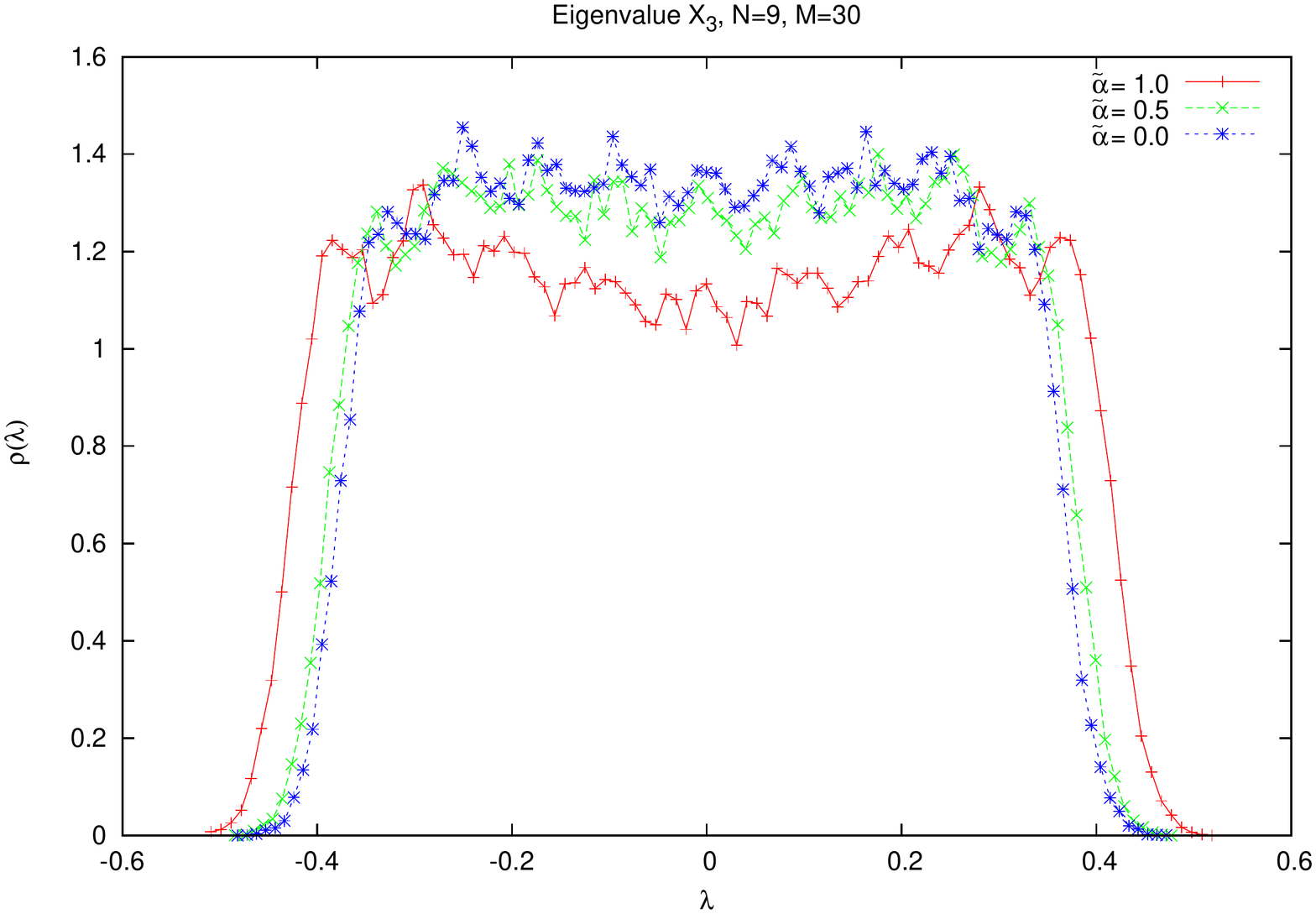}\\
\includegraphics[width=14.0cm,angle=0]{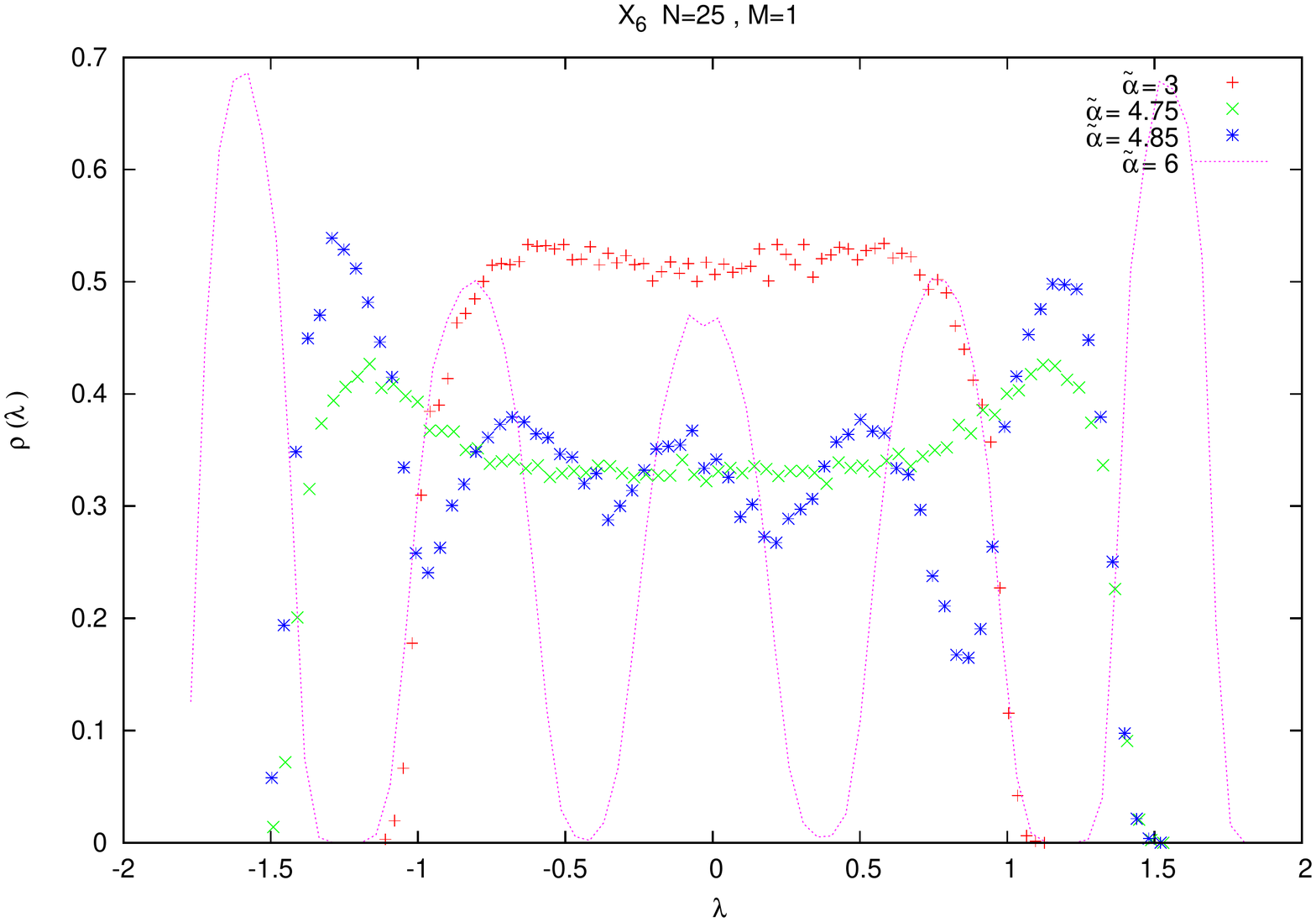}\\
\caption{The eigenvalue distributions for large values of $M$.}\label{ev22}
\end{center}
\end{figure}

\begin{figure}[htbp]
\begin{center}
\includegraphics[width=14.0cm,angle=0]{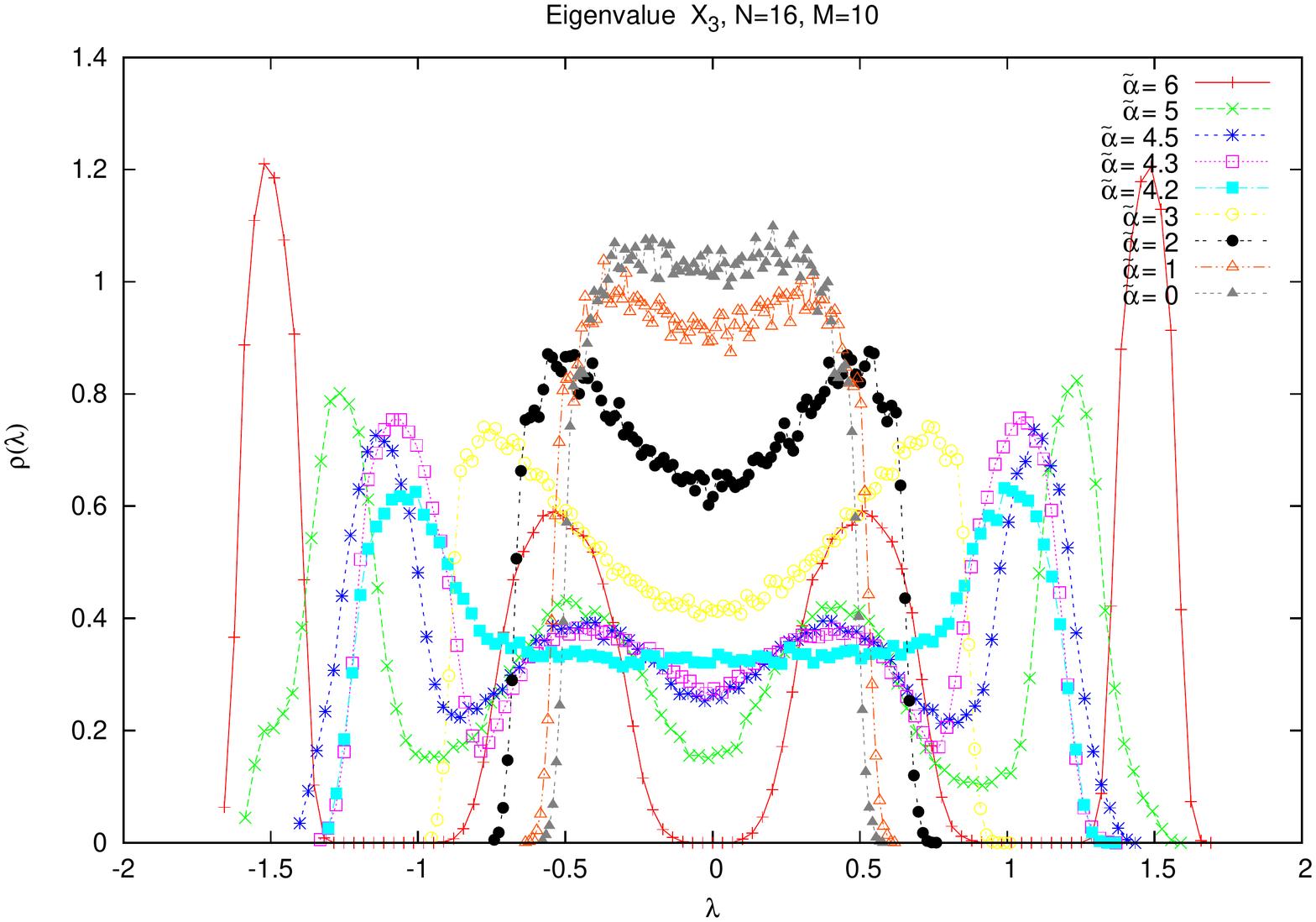}\\
\includegraphics[width=14.0cm,angle=0]{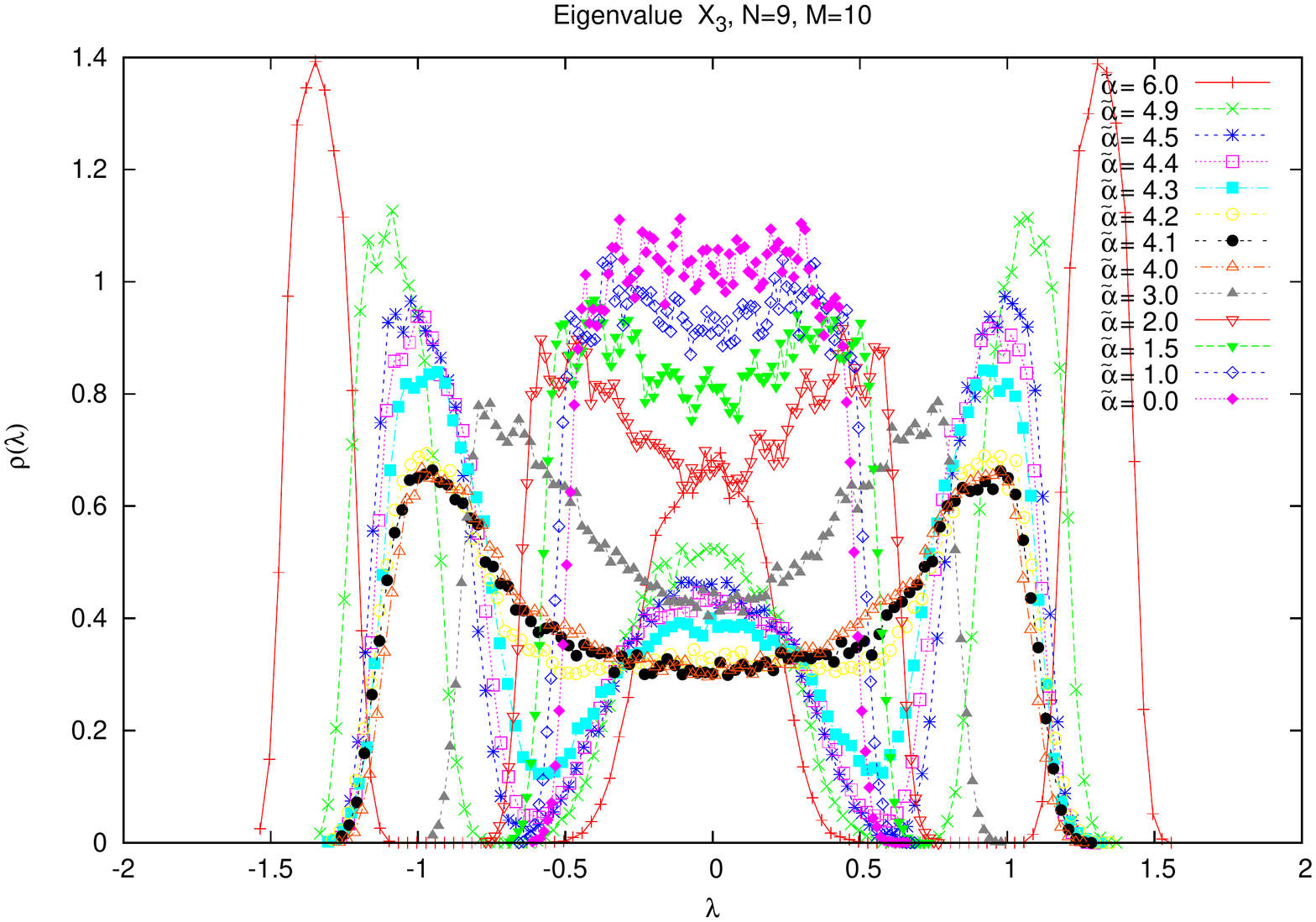}
\caption{The eigenvalue distributions for large values of $M$.}\label{ev22e}
\end{center}
\end{figure}

\begin{figure}[htbp]
\begin{center}
\includegraphics[width=14.0cm,angle=0]{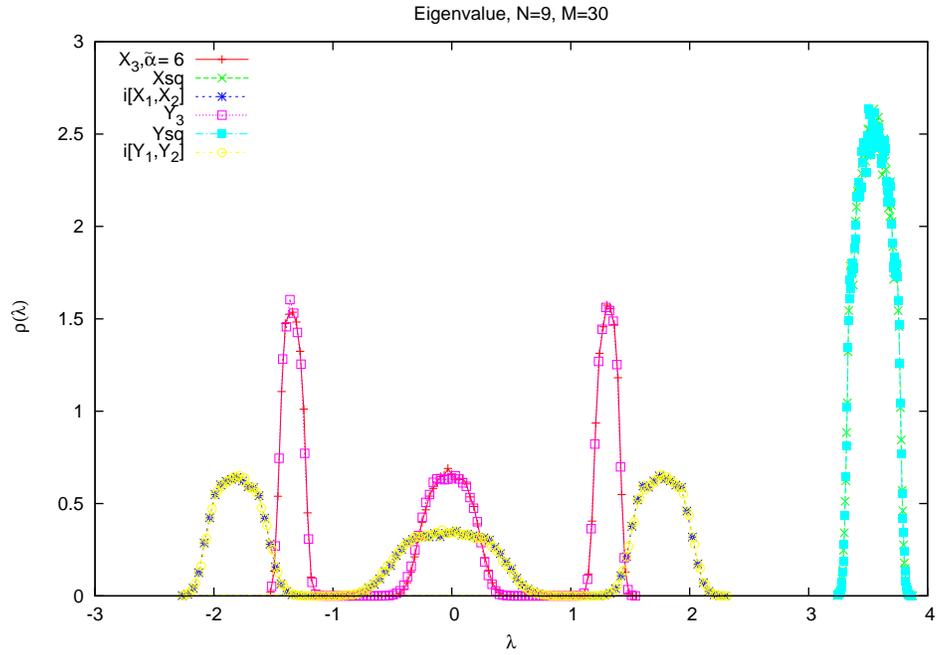}
\includegraphics[width=14.0cm,angle=0]{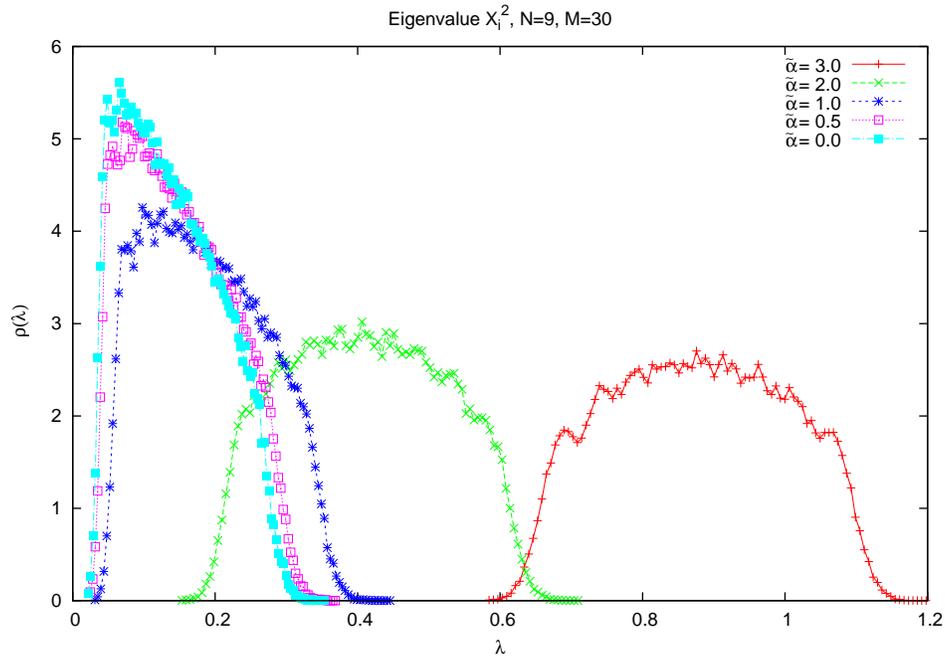}
\caption{Rotational invariance in the eigenvalue distributions for large values of $M$.}\label{ev22ee}
\end{center}
\end{figure}

\begin{figure}[htbp]
\begin{center}
\includegraphics[width=14.0cm,angle=0]{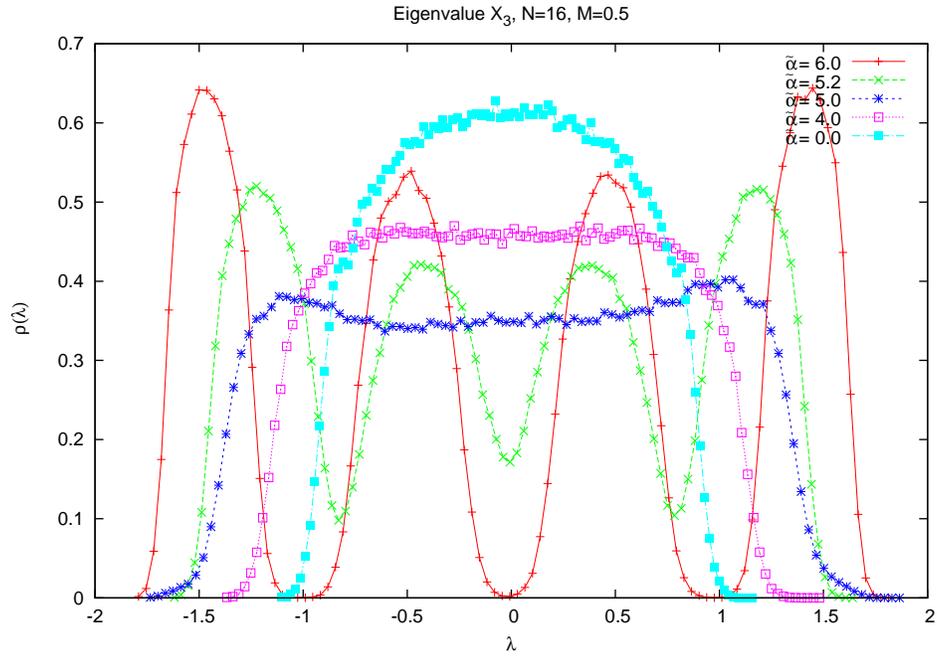}
\includegraphics[width=14.0cm,angle=0]{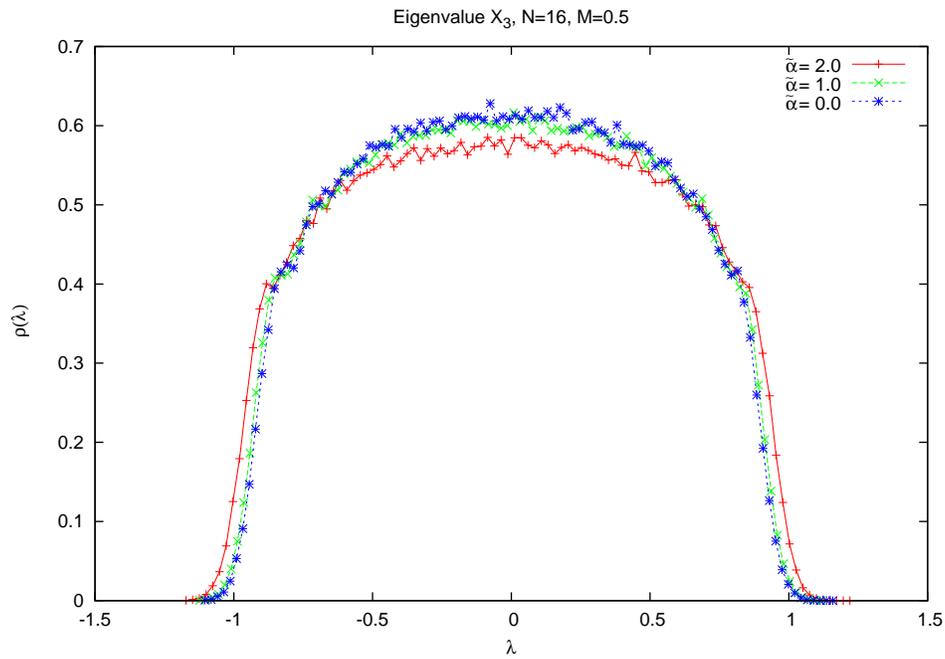}
\caption{The eigenvalue distributions for the Steinacker value $M=1/2$.}\label{ev3}
\end{center}
\end{figure}

\begin{figure}[htbp]
\begin{center}
\includegraphics[width=14cm,angle=0]{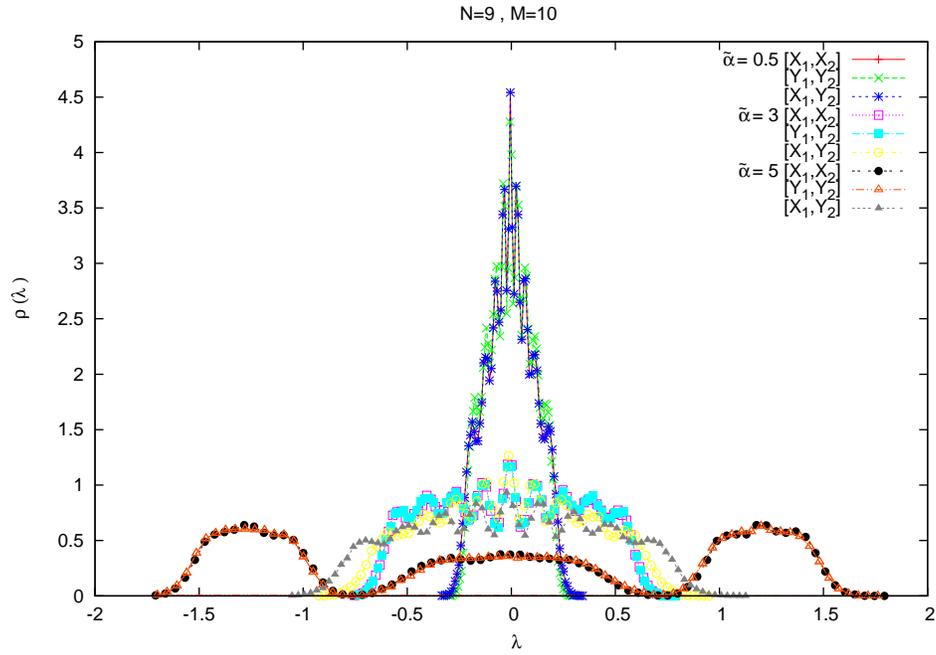}\\
\includegraphics[width=14.0cm,angle=0]{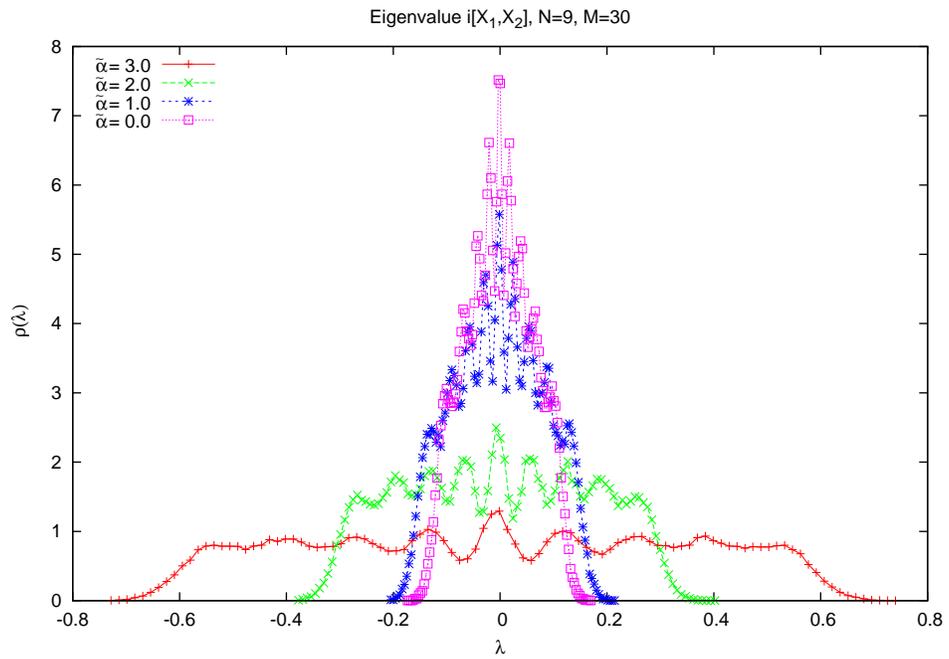}
\end{center}
\caption{The eigenvalue distributions of the commutators $i[X_1,Y_1]$, etc.}\label{ev5}
\end{figure}

\begin{figure}[htbp]
\begin{center}
\includegraphics[width=8cm,angle=0]{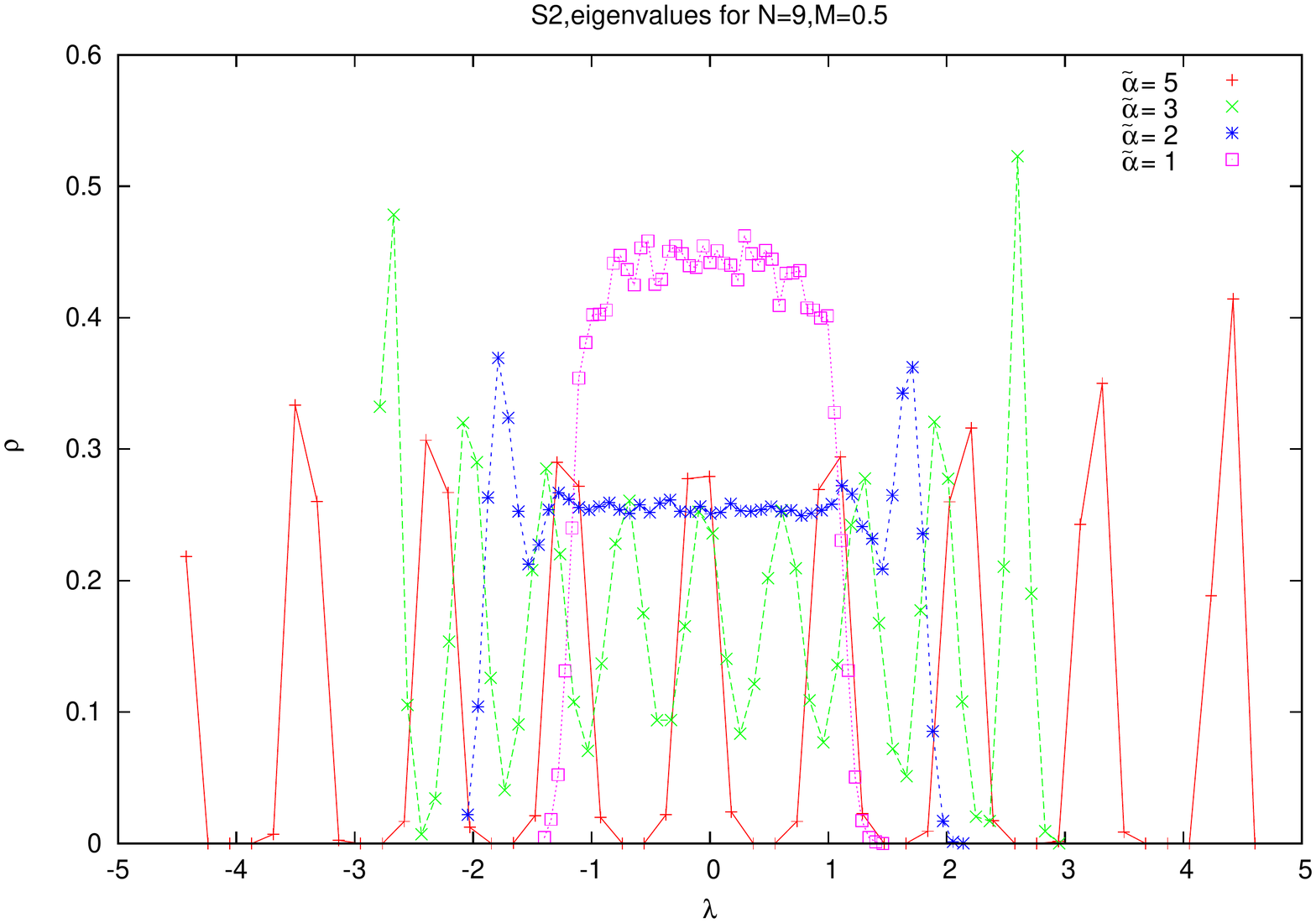}
\includegraphics[width=8cm,angle=0]{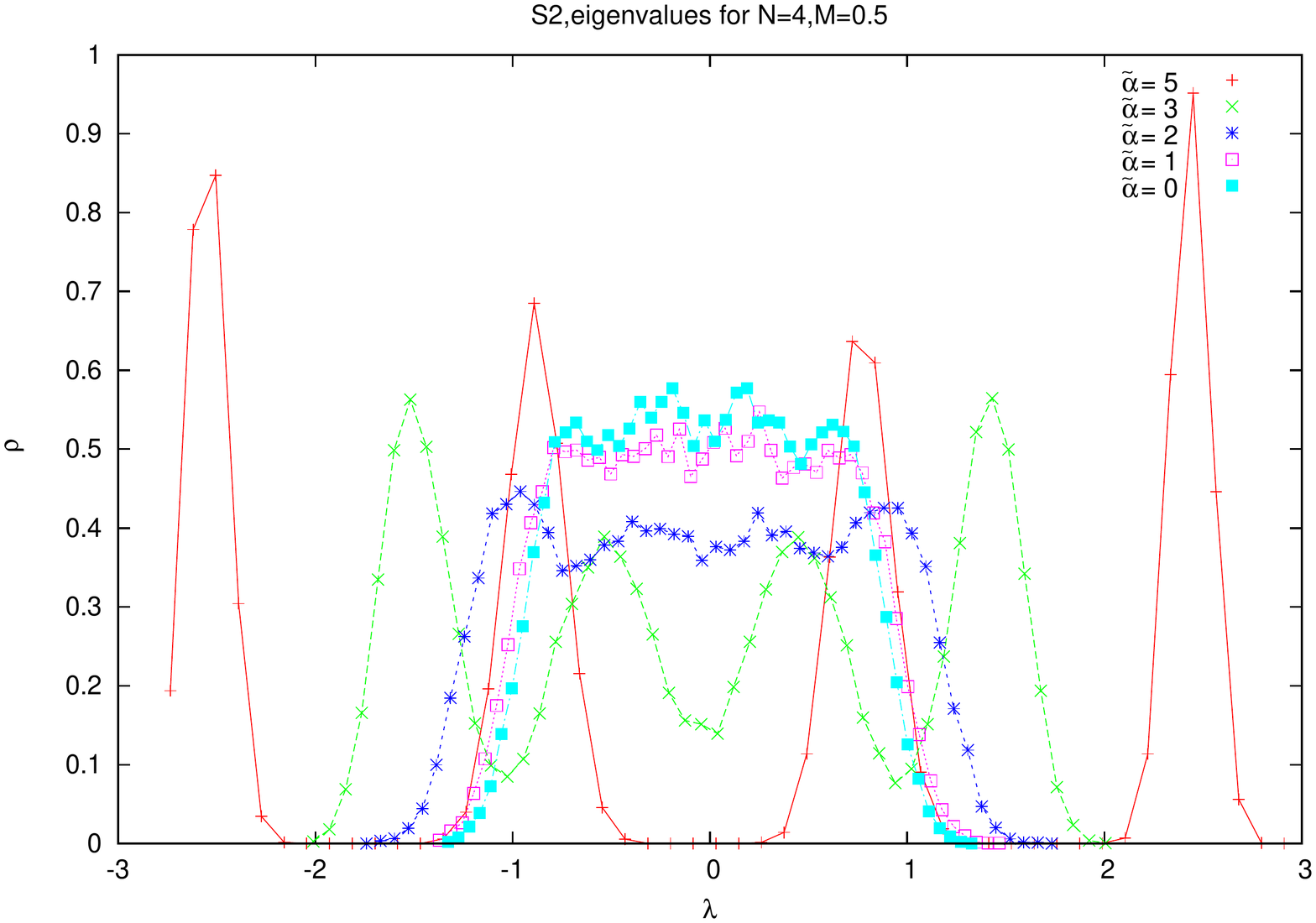}
\includegraphics[width=8cm,angle=0]{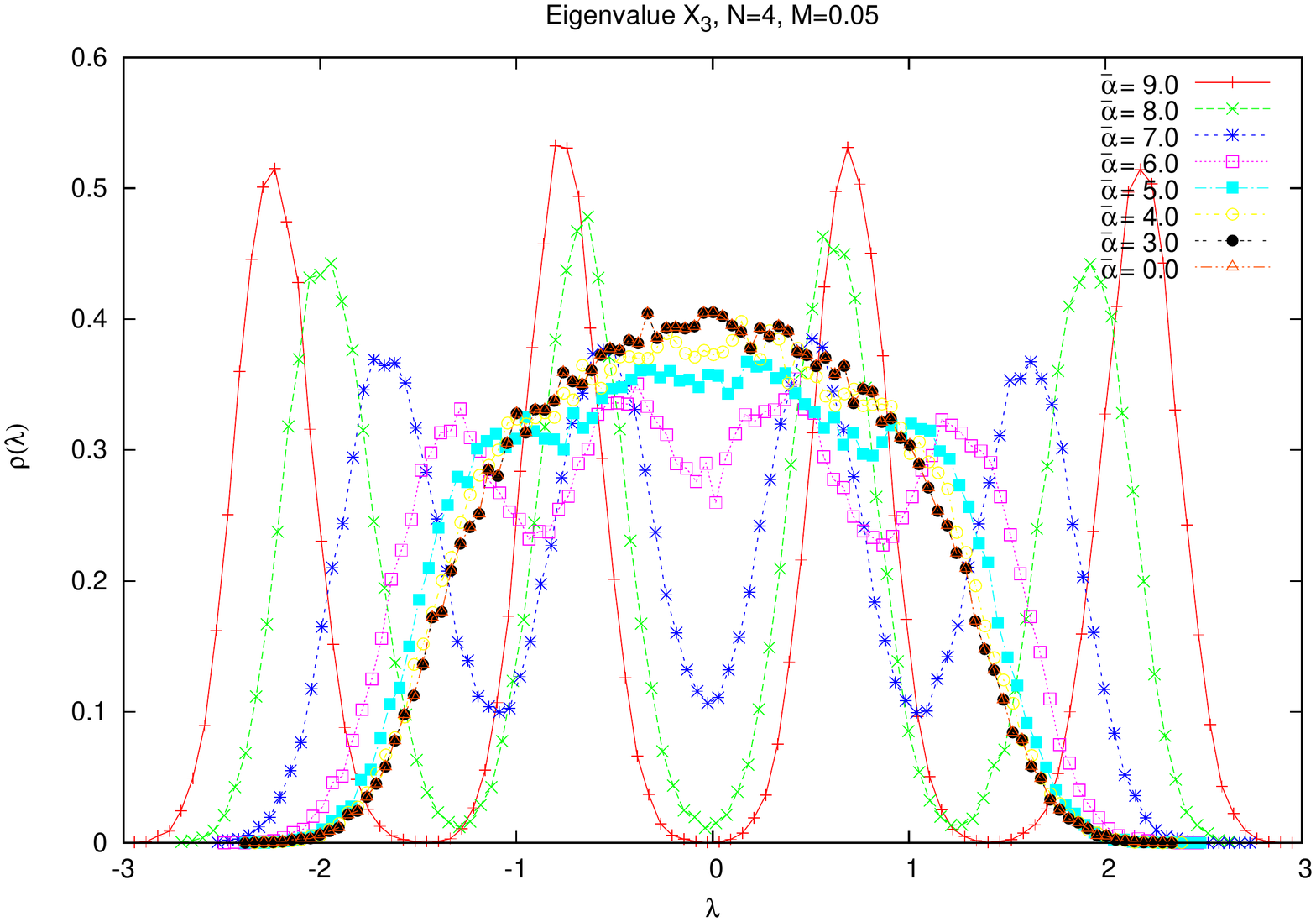}
\includegraphics[width=8cm,angle=0]{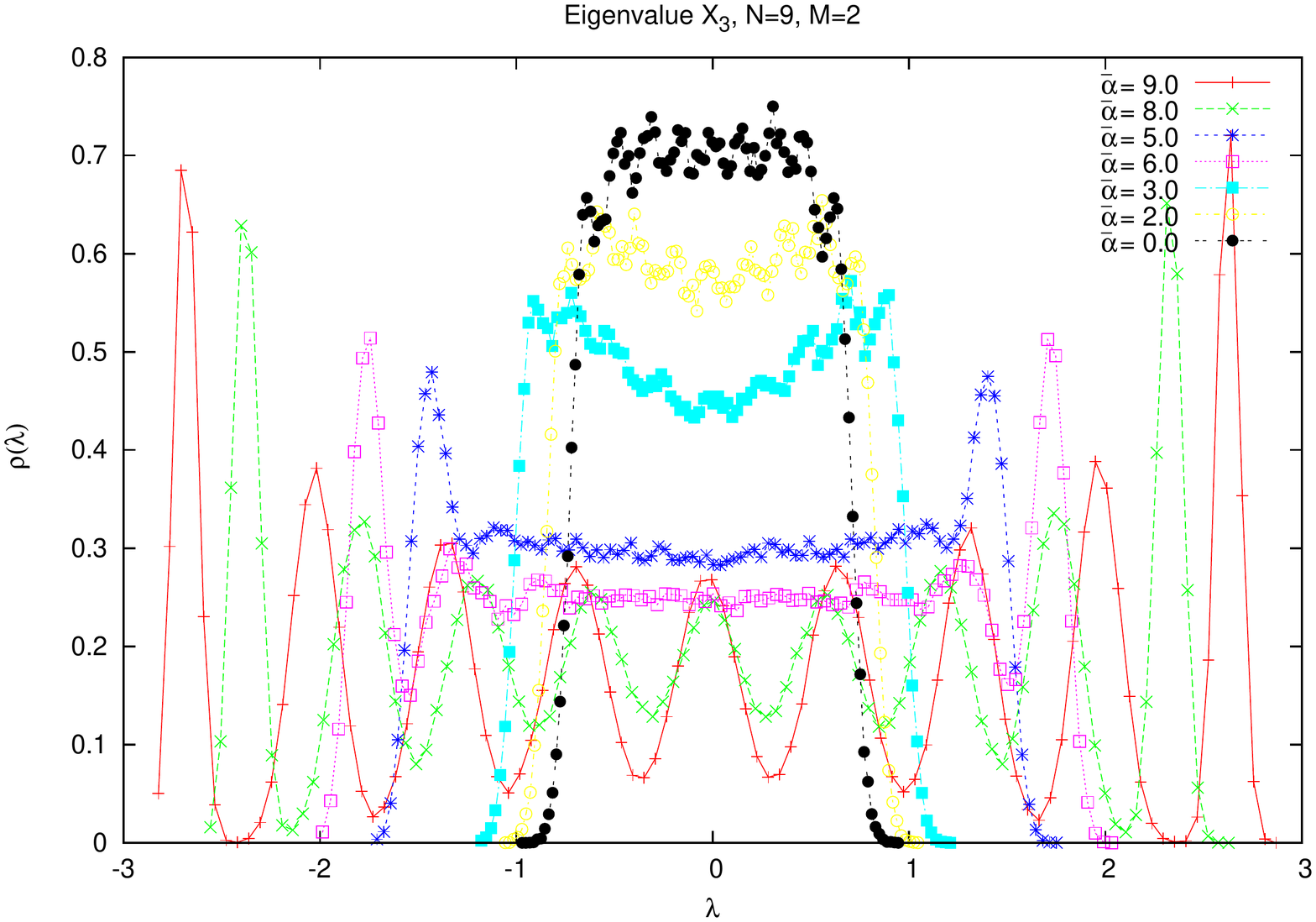}
\end{center}
\caption{The eigenvalue distributions on a single sphere.}\label{ev4}
\end{figure}

\paragraph{Small $M$:} The behavior of the eigenvalue distributions for small values of $M$, such as $M=0.05$ and $M=0.01$, is depicted on figure (\ref{ev1}). The two limiting behaviors are as follows:
\begin{itemize}
\item For large values of the gauge coupling constant we observe a point spectrum given by the eigenvalues of the $SU(2)$ generators in the largest irreducible representation which is of size $N$.
\item Motivated by the work \cite{Berenstein:2008eg} it was conjectured in \cite{Ydri:2012bq} that the joint eigenvalues distribution of $d$ matrices $X_1$, $X_2$,...$X_d$ with dynamics given by a reduced Yang-Mills action should be uniform inside a solid ball of some radius $R$. See also \cite{Filev:2013pza,O'Connor:2012vr,Filev:2014jxa}.  Let $\rho(x_1,...,x_d)$ be the joint eigenvalues distribution of the $d$ matrices $X_1$, $X_2$, ... and $X_d$. We assume that $\rho(x_1,...,x_4)$ is uniform inside a four dimensional ball of radius $r$. The eigenvalues distribution of a single matrix, say $X_d$, which is induced by integrating out the other $d-1$ matrices is given by
\begin{eqnarray}
\rho(\lambda)=\frac{\Omega_{d-1}}{V_d(d-1)}(r^2-\lambda^2)^{(d-1)/2}.\label{d=6}
\end{eqnarray}
\item For small values of the gauge coupling constant $\tilde{\alpha}\longrightarrow 0$ we observe a very good agreement with this law with $d=6$.%given by 
%\begin{eqnarray}
%\rho(\lambda)=\frac{3}{4R^3}(R^2-\lambda^2).
%\end{eqnarray}
%For example, (fit)
%\begin{eqnarray}
%\rho(\lambda)=\frac{\Omega_{d-1}}{V_d(d-1)}(r^2-\lambda^2)^{(d-1)/2}.
%\end{eqnarray}

We also include in the second graph of figure (\ref{ev1}), the $d=4$ law and the $d=3$ parabolic law, for comparison. The fit for $N=16$,  $M=0.01$, $\tilde{\alpha}=0$ gives a value of the radius $r$ of the distribution given by 
\begin{eqnarray}
r\simeq 2.
\end{eqnarray}
We believe that this is a universal behavior at small $\tilde{\alpha}$ and small $M$.  
\end{itemize}
\paragraph{Large $M$:} Some data is included on figures  (\ref{ev22}) and (\ref{ev22e}).
\begin{itemize}
\item For large values of the gauge coupling constant we still observe a point spectrum given by the eigenvalues of the $SU(2)$ generators.

\item For small values of the gauge coupling constant $\tilde{\alpha}\longrightarrow 0$ we observe approximately a uniform distribution. This is neatly shown for $M=30$, $N=9$ and $M=1$, $N=25$ on figure  (\ref{ev22}). The Yang-Mills matrix phase is then characterized, for large values of $M$, by this one-cut uniform distribution as opposed to the $d=6$ law which characterizes the Yang-Mills phase for small values of $M$. The transition from the fuzzy four-sphere to the uniform distribution goes through a new phase or a crossover as we will now discuss.

\item {\bf A new phase or a crossover:} There seems to be another phase appearing for large values of $M$  between the fuzzy four-sphere phase and the Yang-Mills phase. This is indicated by the transition from the distinct point spectrum in the fuzzy four-sphere phase to a phase where a strong gauge field is superimposed on the four-sphere background in such a way that the middle peaks flatten then disappears slowly in favor of the uniform distribution. The last peaks to go are the maximum and the minimum of the $SU(2)$ configuration. See figure (\ref{ev22e}) for $N=16,9$ and $M=10$ where this transition occurs at $\tilde{\alpha}=4.2$ at the peak of $C_v$. Recall that the peak of the specfic heat saturates for large values of $\tilde{\alpha}$ at  $\tilde{\alpha}=4.2$.

However, this new phase may only be a crossover transition. Indeed, on the second graph of (\ref{ev22ee}) we plot the behavior of $X_a^2$ as a function of $\tilde{\alpha}$ for $N=9$ and $M=30$. We observe that the profile of the eigenvalue distribution changes drastically only at the transition point to the uniform distribution.

\item {\bf The coupling between the two spheres:} The coupling between the two spheres can be probed by the eigenvalue distribution of the commutator $i[X_1,Y_1]$. A sample is shown on figure (\ref{ev5}) for $N=9$ and $M=10$. We observe deep inside the fuzzy four-sphere phase that the commutator $i[X_1,Y_1]$ is very different from the rotationally identical commutators $i[X_1,X_2]$ and $i[Y_1,Y_2]$, whereas the three commutators behave indistinguishably from each other deep inside the Yang-Mills matrix phase. 

In the middle phase, during the crossover, the eigenvalue distribution of  $i[X_1,Y_1]$ approaches quickly the profile of the other two. The widths, for example, become less and less different, and shrink to a minimum value in the limit $\tilde{\alpha}\longrightarrow 0$. This is shown explicitly for the commutator  $i[X_1,X_2]$ for $N=9$ and $M=30$.

\item {\bf Steinacker's value $M=0.5$ and transition point in $M$:} See (\ref{ev3}). The behavior of the eigenvalue distribution deep inside the Yang-Mills matrix phase for $M=0.5$, and other medium values of $M$, is neither given by the $d=6$ law, nor it is given by a uniform distribution. This indicates that this value is near the triple point where the transition line between the fuzzy four-sphere ${\bf S}^2\times{\bf S}^2$ and the Yang-Mills matrix phase becomes a crossover phase.

\item {\bf Rotational symmetry:} The eigenvalue distributions of $X_3$, $Y_3$, and the commutators $i[X_1,X_2]$, $i[Y_1,Y_2]$, and the squares $X_a^2$ and $Y_a^2$, in the fuzzy four-sphere phase are shown on the first graph of (\ref{ev22ee}). This shows explicitly the rotational symmetry between the two spheres.
\end{itemize}
\paragraph{The case of a single sphere:} The physics in this case is very similar to the case of the fuzzy four-sphere and a sample of the eigenvalue distributions is included on figure (\ref{ev4}).

%\begin{figure}[htbp]
%\begin{center}
%\includegraphics[width=14.0cm,angle=0]{data_ramda/ev_X_6_N=25_M=1.pdf}\\
%\caption{The eigenvalue distributions for medium values of $M$.}\label{ev2}
%\end{center}
%\end{figure}

\subsection{The effective potential revisited}
We recall the formula for the effective potential%, the equation of motion and the critical values given by
\begin{eqnarray}
\frac{V}{2N^2}%&=&2c_{2}^{0}{\alpha}^4\bigg[\frac{\phi^4}{4}-\frac{{\phi}^3}{3}+m^2\frac{\phi^4}{4}-\mu\frac{\phi^2}{2}\bigg]+\log\phi^2\nonumber\\
&=&\tilde{\alpha}_0^4\bigg[\frac{\phi^4}{4}-\frac{{\phi}^3}{3}+m^2\frac{\phi^4}{4}-\mu\frac{\phi^2}{2}\bigg]+\log\phi^2.
\end{eqnarray}
The classical solutions are given by the condition 
\begin{eqnarray}
\frac{V^{'}}{2N^2}=0.\label{23}
\end{eqnarray}
Clearly the effective potential is not bounded at $\phi=0$. However, this should not pose any problem since this potential is valid for large values of $\tilde{\alpha}$ where we know that the fuzzy sphere exists. For smaller values of $\tilde{\alpha}$ the fuzzy sphere configuration ceases to exist and we enter the matrix phase as we have seen in Monte Carlo data. 

The classical potential admits as solutions $\phi=0$ (local minimum) together with the fuzzy four-sphere solution $\phi_+$ (global minimum),  and a maximum of the barrier between them denoted by $\phi_-$ (local maximum). The solutions $\phi_{\pm}$ exist for $t>-1/4$. As we decrease $\mu$ towards negative values the global minimum $\phi_+$ becomes degenerate with $\phi=0$ at $t=\mu(1+m^2)=-2/9$ and the height at the maximum of the barrier becomes $\tilde{\alpha}^4/324(1+m^2)^3$.  There exists therefore a first order transition in $\mu$, at $\mu=-2/9(1+m^2)$ for every fixed $m^2$,  which is of the same character as the one at $m^2=0$, from a fuzzy-four sphere phase for $\mu>-2/9(1+m^2)$ to $X_a=Y_a=0$ for $\mu<-2/9(1+m^2)$ . For our value $\mu=2(2m^2-1)/9$ the global minimum is always $\phi=2/3$ and it is separated from $\phi=0$ by a barrier.

The effective potential admits four real solutions. The largest positive solution is $\phi_+$ while the other positive solution gives the local maximum and it will determine the height of the barrier in the effective potential. At the critical point these two solutions merge and the barrier disappears. This is different from the classical solution where the barrier never disappears. The solution $\phi_+$ ceases to exist at the critical value determined by the condition  
 \begin{eqnarray}
\frac{V^{''}}{2N^2}=0.\label{32}
\end{eqnarray}
%where we have redefined the coupling constant by 
% \begin{eqnarray}
%\frac{N_0^2}{2}{\alpha}^4=\tilde{\alpha}_0^4.
%\end{eqnarray}
%\begin{eqnarray}
%\frac{V^{'}}{2N^2}
%&=&\tilde{\alpha}_0^4\bigg[\phi^3-{\phi}^2+m^2\phi^3-\mu \phi\bigg]+\frac{2}{\phi}.
%\end{eqnarray}
%The difference between the result on ${\bf S}^2$ and this result lies in the replacement $\tilde{\alpha}\longrightarrow\tilde{\alpha}_0$ and the replacement $c_2\longrightarrow c_2^0$ in the definition of $\mu$. The analysis of the phase structure is therefore identical. 
%In the limit $m^2\longrightarrow \infty$, we have $\mu\longrightarrow m^2$, and we find a critical line separating the fuzzy sphere phase solution with $\phi\neq 0$, from the Yang-Mills matrix phase solution with $\phi=0$, given by the formula
We recall also that the critical value is found by solving (\ref{23}) and (\ref{32}) and is given implicitly by
\begin{eqnarray}
&&(1+m^2){\phi}_*=\frac{3}{8}\big(1+\sqrt{1+\frac{32t}{9}}\big)\nonumber\\
&&\frac{1}{\tilde{\alpha}_{0*}^4}=\frac{\phi_*^2(\phi_*+2\mu)}{8}\nonumber\\
&&t=\mu(1+m^2).
\end{eqnarray}
The critical value is sent to infinity for $\phi_*=-2\mu$ which is equivalent to $t=-1/4$. Thus quantum mechanically the fuzzy four-sphere may exist for  $t>-1/4$ which is to be compared with the classical prediction $t>-2/9$. 

In this region the classical potential is always positive and thus one should consider all $SU(2)$ representations which are degenerate with $X_a=Y_a=0$, i.e. for which $\sum_in_ic_2(n_i)/N\longrightarrow 0$ in the limit $N\longrightarrow 0$ with $\sum_in_i=N$. However, for large $\tilde{\alpha}$ the ground state is dominated by the representation with the smallest Casimir. The fuzzy four-sphere is therefore not stable in this regime and the critical line between the fuzzy four-sphere phase and the Yang-Mills matrix phase asymptotes the line $t=-2/9$ as shown in \cite{DelgadilloBlando:2012xg}.

What interests us the most in this section is the expansion of the solution $\phi$ around the critical value $\phi^*$. This is given in \cite{DelgadilloBlando:2008vi}. The only difference between the result on ${\bf S}^2$ given in \cite{DelgadilloBlando:2008vi} and our result her lies in the replacement 
\begin{eqnarray}
\tilde{\alpha}\longrightarrow\tilde{\alpha}_0
\end{eqnarray}
and the replacement 
\begin{eqnarray}
c_2\longrightarrow c_2^0 
\end{eqnarray}
in the definition of $\mu$. We get the solution 
\begin{eqnarray}
\phi={\phi}_*+
\frac{4}{{\tilde\alpha}_{0*}^{\frac{5}{2}}}\frac{1}{\sqrt{3{\phi}_*+4\mu}}\sqrt{\tilde{\alpha}_0-\tilde{\alpha}_{0*}}+...
\end{eqnarray}
This takes the form
\begin{eqnarray}
\phi={\phi}_*+\sigma~,~
\sigma=\frac{4(2N)^{1/2}}{{\tilde\alpha}_{*}^{\frac{5}{2}}}\frac{1}{\sqrt{3{\phi}_*+4\mu}}\sqrt{\tilde{\alpha}-\tilde{\alpha}_{*}}+...
\end{eqnarray}
In the large $N$ limit we have 
\begin{eqnarray}
\mu\longrightarrow \frac{2NM}{9}.
\end{eqnarray}
\begin{eqnarray}
\phi_*\longrightarrow \sqrt{\frac{\mu}{NM}}=\frac{\sqrt{2}}{3}.
\end{eqnarray}
\begin{eqnarray}
\alpha_*^4\longrightarrow \frac{8N^2M}{\mu^2}=\frac{162}{M}.
\end{eqnarray}
Thus
\begin{eqnarray}
\phi={\phi}_*+\sigma~,~
\sigma=\frac{6}{{\tilde\alpha}_{*}^{\frac{5}{2}}\sqrt{M}}\sqrt{\tilde{\alpha}-\tilde{\alpha}_*}+...
\end{eqnarray}

\subsection{Critical behavior from one-loop effective potential}
Let us start by noting the Schwinger-Dyson identity
 \begin{eqnarray}
&&4<{\rm YM}_1>+4<{\rm YM}_2>+4<{\rm YM}_{12}>+3<{\rm CS}_1>+3<{\rm CS}_2>\nonumber\\
&+&4<{\rm Quar}_1>+4<{\rm Quar}_2>+2<{\rm Quad}_1>+2<{\rm Quad}_2>=6N^2.
\end{eqnarray}
We have the obvious definitions for the various observables 
\begin{eqnarray}
{\rm YM}_1=NTr\bigg[-\frac{1}{4}[X_a,X_b]^2\bigg]~,~{\rm YM}_2=NTr\bigg[-\frac{1}{4}[Y_a,Y_b]^2\bigg]~,~{\rm YM}_{12}=NTr\bigg[-\frac{1}{4}[X_a,Y_b]^2\bigg].
\end{eqnarray}
\begin{eqnarray}
{\rm CS}_1=NTr\bigg[\frac{2i\alpha}{3}{\epsilon}_{abc}X_aX_bX_c\bigg]~,~{\rm CS}_2=NTr\bigg[\frac{2i\alpha}{3}{\epsilon}_{abc}Y_aY_bY_c\bigg].
\end{eqnarray}
\begin{eqnarray}
{\rm Quar}_1=NMTr (X_a^2)^2~,~{\rm Quar}_2=NM Tr(Y_a^2)^2.
\end{eqnarray}
\begin{eqnarray}
{\rm Quad}_1=N\beta Tr X_a^2~,~{\rm Quad}_2=N\beta Tr Y_a^2.
\end{eqnarray}
We obtain immediately a formula for the average action given by
\begin{eqnarray}
\frac{S}{N^2}=\frac{3}{2}+\frac{1}{4N^2}<{\rm CS}_1>+\frac{1}{4N^2}<{\rm CS}_2>+\frac{1}{2N^2}<{\rm Quad}_1>+\frac{1}{2N^2}<{\rm Quad}_2>.
\end{eqnarray}
We compute
\begin{eqnarray}
\frac{1}{4N^2}<{\rm CS}_1>=\frac{1}{4N^2}<{\rm CS}_2>=-\frac{\tilde{\alpha}^4\phi^3}{24 N}.
\end{eqnarray}
\begin{eqnarray}
\frac{1}{2N^2}<{\rm Quad}_1>=\frac{1}{2N^2}<{\rm Quad}_2>=-\frac{\mu\tilde{\alpha}^4\phi^2}{8N}.
\end{eqnarray}
We also note here the formula for the related radius (with $\phi_0=2/3$) 
\begin{eqnarray}
\frac{1}{R}=<\frac{1}{\phi_0^2\tilde{\alpha}^2c_2^0}Tr X_a^2>=<\frac{1}{\phi_0^2\tilde{\alpha}^2c_2^0}Tr Y_a^2>=\big(\frac{3}{2}\big)^2\phi^2.
\end{eqnarray}
Thus
%\begin{eqnarray}
%\frac{S}{N^2}=\frac{3}{2}-\frac{\tilde{\alpha}^4\phi^3}{12 N}-\frac{\mu\tilde{\alpha}^4\phi^2}{4N}.
%\end{eqnarray}
\begin{eqnarray}
\frac{S}{N^2}&=&\frac{3}{2}-\frac{\tilde{\alpha}^4}{4 N}(\frac{1}{3}\phi^3+\mu \phi^2)\nonumber\\
&=&\frac{3}{2}-\frac{\tilde{\alpha}^4}{4 N}(\frac{1}{3}\phi_*^3+\mu \phi_*^2)-\frac{\tilde{\alpha}_*^4}{4 N}(\phi_*^2+2\mu \phi_*)\sigma\nonumber\\
&=&\frac{S_*}{N^2}-\frac{4}{\phi_*}\sigma.\label{ent}
\end{eqnarray}
\begin{eqnarray}
\frac{S_*}{N^2}&=&\bigg(\frac{3}{2}-\frac{4}{3}\big(\frac{\tilde{\alpha}}{\tilde{\alpha}_*}\big)^4-\frac{\tilde{\alpha}^4\mu}{12N}\phi_*^2\bigg)_*\nonumber\\
&=&\frac{3}{2}-\frac{4}{3}\frac{\phi_*+3\mu}{\phi_*+2\mu}\nonumber\\
&\longrightarrow &-\frac{1}{2}.
\end{eqnarray}
This is the value of the average action or entropy in the fuzzy four-sphere phase at the transition point exactly for $M$ large. For $M$ small we get instead the  (constant) value 
\begin{eqnarray}
\frac{S_*}{N^2}
&\longrightarrow &\frac{1}{6}.
\end{eqnarray}
In the Yang-Mills matrix phase we go back to the first line of equation (\ref{ent}) and set $\phi=0$ to get the value $3/2$ at $\tilde{\alpha}=0$. The transition to the Yang-Mills phase occurs quite suddenly for small $M$. Thus the entropy for small $M$ has a discrete jump given by
\begin{eqnarray}
\frac{\Delta S}{N^2}
&\longrightarrow &\frac{4}{3}.
\end{eqnarray}
The average action can also be derived using the formula 
\begin{eqnarray}
\frac{<S>}{N^2}
&=&\frac{3}{2}+\tilde{\alpha}^4\frac{d}{d\tilde{\alpha}^4}\big(\frac{F}{N^2}\big),
\end{eqnarray}
where (with $X_a=\alpha D_a$, $\tilde{\alpha}^4\hat{S}=S$)
\begin{eqnarray}
Z&=&\exp(-F)=\int dD_a \exp\big(\frac{3N^2}{2}\ln\tilde{\alpha}^4-\tilde{\alpha}^4 \hat{S}\big).
\end{eqnarray}
The free energy is given by
\begin{eqnarray}
\frac{F}{2N^2}=\frac{V}{2N^2}+\frac{1}{2}\ln\tilde{\alpha}^4+{\rm constant}.
\end{eqnarray}
Then we can compute the specific heat by the formula 
\begin{eqnarray}
C_v&=&\frac{<S^2>-<S>^2}{N^2}\nonumber\\
&=&\frac{<S>}{N^2}-\tilde{\alpha}^4\frac{d}{d\tilde{\alpha}^4}\big(\frac{<S>}{N^2}\big).
\end{eqnarray}
We compute immediately 
\begin{eqnarray}
C_v&=&\frac{3}{2}+\frac{\tilde{\alpha}^5}{16 N}\phi(\phi+2\mu)\frac{d\phi}{d\tilde{\alpha}}.%\nonumber\\
%&=&\frac{3}{2}+\frac{1}{4(2N)^{1/2}}\frac{\phi_*(\phi_*+2\mu)}{\sqrt{3\phi_*+4\mu}}\frac{\tilde{\alpha}_*^{5/2}}{\sqrt{\tilde{\alpha}-\tilde{\alpha}_*}}.
\end{eqnarray}
From this equation we can derive immediately the divergent part of the specific heat to be given by
\begin{eqnarray}
C_v
&=&C_v^B+\frac{1}{4(2N)^{1/2}}\frac{\phi_*(\phi_*+2\mu)}{\sqrt{3\phi_*+4\mu}}\frac{\tilde{\alpha}_*^{5/2}}{\sqrt{\tilde{\alpha}-\tilde{\alpha}_*}}\nonumber\\
&=&C_v^B+\frac{3^{1/2}}{2^{7/8}M^{1/8}}\frac{1}{\sqrt{\tilde{\alpha}-\tilde{\alpha}_*}}.
\end{eqnarray}
The critical exponent of the specific heat is $1/2$ which is precisely the value obtained in two dimensions. Similarly, the critical value $\tilde{\alpha}_*$ and the coefficient of the singularity become vanishingly small when $M\longrightarrow \infty$.

The behavior of the background specific heat $C_v^B$ deep inside the fuzzy four-sphere phase is computed as follows. The solution $\phi$ of the equation of motion for large values of $\tilde{\alpha}$ is found to be given by 
\begin{eqnarray}
\phi=\frac{2}{3}-\frac{27}{2M\tilde{\alpha}^4}+....
\end{eqnarray}
By substitution we get 
\begin{eqnarray}
C_v^B
&=&\frac{3}{2}+\frac{\tilde{\alpha}^5}{16 N}\phi(\phi+2\mu)\frac{54}{M\tilde{\alpha}^5}\nonumber\\
&=&\frac{3}{2}+1\nonumber\\
&=&\frac{5}{2}.
\end{eqnarray}
\section{Related topics}
\paragraph{Emergent gauge theory in two dimensions:}
We have established that the fuzzy sphere is completely stable in the three matrix model (\ref{fuzzys2}) for any value of $M$. This includes Steinacker's value $M=1/2$. We can now speak in a consistent way about constructing $U(n)$ gauge theory on fuzzy ${\bf S}^2$. This entails the construction of monopoles sectors and the direct evaluation of the partition function on the fuzzy sphere, or alternatively its evaluation by means of localization technique, as a sum over instantons contributions. This has been done for Steinacker's value in \cite{Steinacker:2003sd} and \cite{Steinacker:2007iq} respectively. The result in this case was found to be identical to the result on the ordinary sphere. This shows that the fuzzy sphere is acting here as a regularized version of the sphere, and also shows that this matrix method is potentially a powerful new method for gauge theory. 

The construction of fuzzy monopoles and instantons and the corresponding Ginsparg-Wilson fermions on this stable fuzzy sphere can also be carried out along the lines of \cite{Balachandran:2003ay,Balachandran:2004dj,Balachandran:2002jq}.

\paragraph{A stable four-sphere ${\bf S}^2\times{\bf S}^2$ and topology change:}
The main conclusion we have reached in this article is the difference between fuzzy ${\bf S}_N^2$ and fuzzy ${\bf S}^2_N\times{\bf S}^2_N$. The fuzzy four-sphere is only stable in the large $M$ limit. A simple modification of the six matrix model may lead to a stable four-dimensional geometry for any $M$. This consists in adding the terms 
\begin{eqnarray}
N \beta_1( Tr X_a^2+ Tr Y_a^2),
\end{eqnarray}
with a particular coefficient $\beta_1$. This will modify the critical line (\ref{theo}) in such a way that $\tilde{\alpha}$ is replaced by $\bar{\alpha}$.

Topology change can also be easily obtained in the above six matrix model by setting zero the mass parameters $M$ and $\beta$ on one of the spheres. This way we will have the possibility of a transition between the fuzzy four-sphere  ${\bf S}^2_N\times{\bf S}^2_N$ phase and the fuzzy sphere  ${\bf S}^2_N$ phase. This is what we call a topology change. This scenario was previously observed on fuzzy  ${\bf CP}^n$   \cite{Dou:2007in}.

\paragraph{Critical behavior:} The critical behavior of the above stable fuzzy two-sphere ${\bf S}^2_N$ and also the critical behavior of the fuzzy four-sphere ${\bf S}^2_N\times{\bf S}^2_N$ can also be determined more carefully along the line of  \cite{O'Connor:2013rla}.

\paragraph{Comparison with \cite{Behr:2005wp}, the $2-$matrix model and instanton calculus:} The six matrix model (\ref{action00}) with $M=1/2$ is precisely the action considered in  \cite{Behr:2005wp}. 

Since this model is only stable for large $M$, the corresponding instanton calculus should only be expected to be valid deep inside the fuzzy four-sphere phase.

A $2-$matrix model associated with the $6-$dimensional Yang-Mills matrix model (\ref{action00}) can also be constructed starting from an $SO(6)$ formulation. This should be contrasted with the  $SO(3)$ formulation on the fuzzy sphere which leads to (\ref{ba}). In the case of the fuzzy four-sphere ${\bf S}^2_N\times{\bf S}^2_N$ the matrices $C$ and $D$ corresponding to $X_a$ and $Y_a$ respectively are found to be highly constrained  \cite{Behr:2005wp} as opposed to the single constraint $C_0=0$ found on the fuzzy sphere.

The six matrix model considered in this article is also closely related to the action studied in \cite{Imai:2003ja}.
\paragraph{Dirac operator:} There are two seemingly different formulations of the Dirac operator on the fuzzy four-sphere ${\bf S}^2_N\times{\bf S}^2_N$. The one presented in  \cite{Behr:2005wp} is again based on $SO(6)$ whereas the one presented in \cite{bal_private} is based on $SO(3)\times SO(3)$.

\paragraph{Fuzzy four-sphere $S^4$ and fuzzy ${\bf CP}^2$:}
The above analysis is expected to hold without much change on fuzzy ${\bf S}^4$ \cite{Medina:2002pc,Medina:2012cs,Steinacker:2015dra} and fuzzy ${\bf CP}^2$ \cite{Grosse:1999ci,Grosse:2004wm,Azuma:2004qe,Alexanian:2001qj}. The analogous actions on fuzzy  ${\bf CP}^n$ are given by  \cite{Dou:2007in}
\begin{eqnarray}
S
&=&\frac{1}{g^2N}Tr\bigg[-\frac{1}{4}[D_a,D_b]^2+i\epsilon_{abc}D_aD_bD_c\bigg]+\frac{3n}{4g^2N}Tr\Phi +\frac{M_0^2}{N}  Tr{\Phi}^2+ \frac{M^2}{N}\frac{(2n+3)^2}{36n} Tr{\Phi}.\nonumber\\
\end{eqnarray}
\section{Conclusion}

In this article we have studied IKKT Yang-Mills matrix models with mass deformations in three and six dimensions.

The $3-$dimensional IKKT matrix models considered here are very similar to the ones studied in \cite{CastroVillarreal:2004vh,O'Connor:2006wv}. However, the dynamically emergent geometry, which is given by a fuzzy two-sphere ${\bf S}^2_N$,  is found to be stable for all values of the deformation parameter $M$. This was anticipated previously for $M=1/2$ in \cite{Steinacker:2003sd}. Indeed, the critical gauge coupling constant $\tilde{\alpha}$ is found to scale as in equation (\ref{stable_sphere}), i.e. as $\tilde{\alpha}\sim 1/\sqrt{N}$. The sphere-to-matrix transition line is pushed to 0 and only one phase survives.

 In this case the fuzzy sphere acts then as a regulator of the commutative sphere, and as a consequence, fuzzy field theory and fuzzy physics, based on this emergent fuzzy sphere, makes full sense for all values of the gauge coupling constant.

We have also  studied in this article $6-$dimensional IKKT matrix models, with global $SO(3)\times SO(3)$ symmetry, containing
at most quartic powers of the matrices proposed in \cite{DelgadilloBlando:2006dp}. The value $M=1/2$ of the deformation corresponds to the model of \cite{Behr:2005wp}. This theory exhibits a phase transition from a geometrical phase at low temperature, given by a fuzzy four-sphere ${\bf S}^2_N\times{\bf S}^2_N$ background, to a Yang-Mills  matrix phase with no background geometrical structure at high temperature. 

The geometry as well as an Abelian gauge field and two scalar fields are determined dynamically as the temperature is decreases and the fuzzy four-sphere condenses. 

The transition is exotic in the sense that we observe, for small values of $M$, a discontinuous jump in the entropy, characteristic of a 1st order
transition, yet with divergent critical fluctuations and a divergent
specific heat with critical exponent $\alpha=1/2$. The critical temperature is pushed upwards as the 
scalar field mass is increased.  For small $M$, the system in the Yang-Mills phase is well
approximated by $6$ decoupled matrices with a joint eigenvalue distribution which is uniform inside a ball in ${\bf R}^6$. This gives what we call the $d=6$ law given by equation (\ref{d=6}). For large $M$, the transition from the four-sphere phase to the Yang-Mills matrix phase turns into a crossover and the eigenvalue distribution in the Yang-Mills matrix phase changes from the $d=6$ law to a uniform distribution. 

 In the Yang-Mills matrix phase the specific heat is equal to $3/2$ which coincides with the specific heat of $6$ independent matrix models with quartic potential in the high temperature limit  and is therefore consistent with this interpretation. Once the geometrical phase is well established the specific heat 
takes the value $5/2$ with the gauge field contributing $1/2$ and the two scalar fields each contributing $1$.

The $6-$dimensional IKKT Yang-Mills matrix models studied here present thus an
appealing picture of a $4-$dimensional geometrical phase emerging as the system cools
and suggests a scenario for the emergence of geometry in the early
universe. See \cite{Ydri:2016kua} and references therein.

%\newpage
%\begin{figure}

\paragraph{Acknowledgment:}  
This research was supported by CNEPRU: "The National (Algerian) Commission for the Evaluation of
University Research Projects"  under contract number ${\rm DO} 11 20 13 00 09$.

\bibliographystyle{unsrt}

%\end{figure}
\end{document}